\documentclass[twocolumn]{aastex701}

\usepackage{graphics,graphicx}
\usepackage{times}

\usepackage[english]{babel} 
\usepackage[T1]{fontenc} 
\usepackage{lmodern} 
\usepackage{float} 
\usepackage{amsmath} 
\usepackage{amssymb} 

\usepackage{hyperref}

\usepackage{calrsfs,euscript,mathrsfs,amssymb}
\usepackage{textcomp, gensymb}

\newcommand{\nicer}{\textit{NICER} }
\newcommand{\nustar}{{\it NuSTAR} }
\newcommand{\fermi}{{\it Fermi}-LAT }
\newcommand{\amego}{{\it AMEGO} }
\newcommand{\chandra}{{\it Chandra} }
\newcommand{\swift}{{\it Swift}--XRT }

\usepackage{lineno}[mathlines]
\usepackage{breqn}

\begin{document}

\title{A multi-wavelength view of the multi-messenger sources NGC 1068 and PKS 1502+106}

\correspondingauthor{Abhishek Desai}
\email{adesai.physics@gmail.com}

\author[0000-0001-7405-9994]{Abhishek Desai}
\affiliation{NASA Postdoctoral Program Fellow, NASA Goddard Space Flight Center, Greenbelt, MD, 20771, USA}
\email{}

\author[0000-0001-5544-0749]{Stefano Marchesi}
\affiliation{Dipartimento di Fisica e Astronomia (DIFA), Università di Bologna, via Gobetti 93/2, I-40129 Bologna, Italy}
\affiliation{Department of Physics and Astronomy, Clemson University, Kinard Lab of Physics, Clemson, SC 29634-0978, USA}
\affiliation{INAF – Osservatorio di Astrofisica e Scienza dello Spazio di Bologna, Via Piero Gobetti 93/3, 40129 Bologna, Italy}
\email{}

\author[0000-0002-9867-6548]{Justin Vandenbroucke}
\affiliation{Dept. of Physics and Wisconsin IceCube Particle Astrophysics Center, University of Wisconsin{\textendash}Madison, Madison, WI 53706, USA}
\email{}

\author[0000-0002-7825-1526]{Indrani Pal}
\affiliation{Department of Physics and Astronomy, Clemson University, Kinard Lab of Physics, Clemson, SC 29634-0978, USA}
\email{}

\author[0000-0002-5387-8138]{Ke Fang}
\affiliation{Dept. of Physics and Wisconsin IceCube Particle Astrophysics Center, University of Wisconsin{\textendash}Madison, Madison, WI 53706, USA}
\email{}

\author[0000-0002-8028-0991]{Dieter Hartmann}
\affiliation{Department of Physics and Astronomy, Clemson University, Kinard Lab of Physics, Clemson, SC 29634-0978, USA}
\email{}

\author[0000-0002-9280-836X]{Regina Caputo}
\affiliation{NASA Goddard Space Flight Center, Maryland, U.S.A.}
\email{}

\author[0000-0002-6584-1703]{Marco Ajello}
\affiliation{Department of Physics and Astronomy, Clemson University, Kinard Lab of Physics, Clemson, SC 29634-0978, USA}
\email{}

\author[0000-0001-9179-3760]{Jessie Thwaites}
\affiliation{Dept. of Physics and Wisconsin IceCube Particle Astrophysics Center, University of Wisconsin{\textendash}Madison, Madison, WI 53706, USA}
\email{}

\author[0000-0002-2068-6949]{Kavic R Kumar}
\affiliation{University of Maryland, College Park, Maryland, U.S.A}\affiliation{NASA Goddard Space Flight Center, Maryland, U.S.A.}
\email{}

\author[0009-0007-2644-5955]{Sam Hori}
\affiliation{Dept. of Physics and Wisconsin IceCube Particle Astrophysics Center, University of Wisconsin{\textendash}Madison, Madison, WI 53706, USA}
\email{}

\begin{abstract}
Multi-messenger astronomy offers a powerful approach to studying high-energy radiative processes in astrophysical sources. A notable example was seen in 2017, when the IceCube Neutrino Observatory detected a high-energy neutrino event that was found to coincide with a gamma-ray flare from a blazar. Since then, numerous multi-messenger studies combining neutrino and photon data have been conducted, yet the origin of neutrinos from active galactic nuclei (AGN) remains uncertain. In this work, we present the results of an X-ray observing program targeting two AGNs, NGC 1068 and PKS 1502+106. The multi-wavelength dataset includes new observations from \textit{NICER} and \textit{NuSTAR} from the observing proposal along with gamma-ray data collected using \textit{Fermi}-LAT, and one archival observation from \textit{Chandra}. Additionally, we  derive the neutrino fluxes for both AGNs using ten years of IceCube data and neutrino spectra predicted by theoretical models. These results demonstrate the value of combining multi-messenger data in building and constraining theoretical models. They also highlight the importance of testing model predictions against observational data to refine measurements of both the neutrino flux and spectral shape.
\end{abstract}

\submitjournal{ApJ}

%
%

\section{Introduction}
\label{sec:intro}

Active Galactic Nuclei (AGN) are among the most promising candidates for explaining the extragalactic neutrinos observed by the IceCube Neutrino Observatory at the South Pole. However, the exact nature of the processes that lead to the creation of these neutrinos remain unclear. 
The first AGN identified as a neutrino source candidate, TXS 0506+056, is a flaring blazar. It was discovered through a multi-wavelength campaign \citep{TXS_Icecube_fermi_others_multimessenger} and supported by a retrospective identification of a neutrino burst from the same source in 2014-2015 \citep{TXS_Icecube}. 
Using ten years of IceCube data, time-integrated searches identified a Seyfert-II galaxy NGC 1068, as a potential neutrino source  at a significance of $4.2\sigma$ \citep{ngc1068_22}. \cite{10yrPStracks_tessa} report three additional neutrino source candidate blazars, namely TXS 0506+056, PKS 1424+240 and GB6 J1542+6129. More recently, \cite{icecube_4151_hans} found two more non-jetted AGN, namely NGC 4151 and CGCG 420-015, which displayed an excess of neutrinos (2.7$\sigma$ and 2.8$\sigma$ respectively) with respect to background signals. A separate analysis \citep{ngc4151_sreetama} also found an excess of neutrinos in the direction of NGC 4151 at a significance of 2.9$\sigma$.

The sources that have been associated with neutrino detections belong to different AGN classes. Blazars are considerably more luminous in the gamma-ray regime compared to Seyfert AGN due to different viewing angles with respect to the observer. Several theoretical models have been developed to understand the neutrino production in AGNs \citep[see e.g.][]{Rodrigues_2021, Fang:2023vdg, murase_stecker_agn_review}.

In jetted AGN models, high-energy neutrinos are produced by the photomeson production process between relativistic protons accelerated by the jet and target photons inside the jet or external radiation fields. The same protons may also produce gamma-ray emission through proton-induced electromagnetic cascades and proton synchrotron radiation. Such models predict a correlation of neutrino flux with the X-ray as well as gamma-ray photons. Such models were invoked to explain the observations of TXS~0506+056 (see for example \citealt{Zhang:2019dob}) as well as PKS 1502+106 \citep{Rodrigues_2021,Rodrigues24_models}. The blazar PKS 1502+106 was found to be flaring in the radio regime during the time of a neutrino event originating from the same direction \citep{pks_icecube_gcn}. Theoretical models  \citep{Rodrigues_2021,Rodrigues24_models} suggest that the shape of the spectral energy distribution (SED) of the blazar changes when the blazar goes from quiescent to flaring state. This change has been observed across the multi-wavelength photon regime while also affecting the neutrino emission.

The neutrino production theory is different for non-jetted AGNs, which are generally classified as unobscured or obscured depending on their measured Hydrogen line-of-sight column density being below or above the 10$^{22}$ cm$^{-2}$ threshold, respectively. For these AGN, high-energy neutrinos may be produced in various regions of the source, such as the AGN cores and starburst regions. 
Given that the observed neutrino flux from NGC~1068 and other Seyfert galaxies significantly exceeds the gamma-ray flux, models inspired by IceCube observations have focused on the coronal region as the site of neutrino production. In these models, protons are accelerated by stochastic processes in turbulent magnetic fields \citep{PhysRevLett.125.011101} or by magnetic reconnection \citep{Fiorillo:2023dts}. They interact with X-ray photons from coronal electrons and with nearby thermal protons, producing charged pions that decay into neutrinos. Proton interactions with X-ray photons also generate gamma rays via the Bethe-Heitler process, which then cascade down to MeV energies \citep{Inoue:2019yfs, PhysRevLett.125.011101}. Such coronal models indicate a correlation between the neutrino and X-ray fluxes \citep{KhateeZathul:2024tgu}.

For both of the above cases, a possible correlation between X-rays and neutrinos exists. We aim to study this correlation by systematic analysis of \nicer X-ray observations of two individual AGN sources along with the data collected by IceCube.
We also add gamma-ray observations from \textit{Fermi}-LAT to test the gamma-ray connection. For the purpose of this study we require both of the AGN to be possible or confirmed neutrino sources with relevant archival theoretical models explaining the multi-messenger emission. We thus select the Seyfert galaxy NGC 1068, detected by \cite{ngc1068_22} and modeled by \cite{ngc1068_theory_model,Murase_2022_hidden_heart_AGN} and the blazar PKS 1502+106 reported as a possible neutrino source by \cite{pks_icecube_gcn} and modeled by \cite{Rodrigues_2021,Rodrigues24_models}.

%
%

\section{X-ray observations:}
\label{sec:x_ray_data_sed}
We report in this section the properties of the monitoring campaign of PKS 1502+106 and NGC 1068 using the \nicer telescope \citep{nicer_orig} for the Guest Observing Cycle 5 run (proposal ID: 6196, PI:Vandenbroucke). We also discuss the joint \nustar observations that were taken as part of this campaign, as well as additional archival observations of the two sources that were taken with other X-ray facilities during the time of the \nicer monitoring. 

Our monitoring campaigns led to a total of 30 \nicer observations (for PKS 1502+106 and NGC 1068) and three joint \nustar observations (for PKS 1502+106).
Additionally, archival data taken by \chandra on January {fourth}, 2024 {(MJD:60313.73)} {were} also used, as it coincided with our \nicer observations. More details of the X-ray data used and the analysis performed are given below.

%
%
%
%

\subsection{\nicer observations}
\label{sec:nicer_x_ray_data_sed}

Out of a total of 30 \nicer observations, we obtained 21 observations for PKS 1502+106 with an observing time of $\sim$7\,ks (spread over 3 to 4 days), and 9 observations for NGC 1068 with an observing time of $\sim$1\,ks for each observation. Two observations of PKS 1502+106 and one observation of NGC 1068 had zero exposure so they were not included in our analysis. 
The {other} 27 non-zero exposure observations were reduced following the standard \nicer data reduction technique and data analysis software (NICERDAS v012).
The reduction and calibration of the event files were performed using the standard \texttt{nicerl2} and \texttt{nicerl3-spec} scripts\footnote{https://heasarc.gsfc.nasa.gov/docs/nicer/analysis_threads/} using the response file from CALDB v.20240206.

For all \nicer observations, no image stacking was performed, as the signal-to-noise, (S/N) ratio for all the source observations was high. 
Also, for each \nicer observation, the source signal was compared with the background signal estimated using the \texttt{3c50} model given by \cite{nicer_3c50_citation}. 
Using this comparison, the energy range for which the source signal was dominating and the S/N ratio was the highest was selected. A fit was then performed using the \texttt{pgstat} statistical method along with a spectral model. The results of the fit for both the sources, NGC 1068 and PKS 1502+106, are described in Tables~\ref{tab:fitting_results_NGC} and ~\ref{tab:fitting_results_PKS}, respectively. For each observation, the energy range used for the individual fits is also  reported in the Tables.

For PKS 1502+106, we use a simple power-law model with Galactic absorption, which is:

\begin{dmath}
    TBabs*powerlaw
\end{dmath}

For NGC 1068, we add 4 additional Gaussian profiles to the model, giving:

\begin{dmath}
    TBabs*(powerlaw+zgauss+zgauss+zgauss+zgauss)
\end{dmath}

The TBabs models the Galactic absorption with fixed values of $N_H = 2.03 \times 10^{20}$ cm$^{-2}$ for PKS 1502+106 and $N_H = 3.32 \times 10^{20}$ cm$^{-2}$ for NGC 1068.
The Gaussian lines are added for NGC 1068, only  when statistically required by the data, to model emission lines seen due to either the source or the \nicer instrument. The Gaussian lines fall around $1.8-2.0\,\mathrm{keV}, 0.8-0.9\,\mathrm{keV}, 0.5-0.6\,\mathrm{keV}$ and $1.2-1.3\,\mathrm{keV}$. While in this initial test a simple power-law model where additional Gaussian lines is used, a more detailed model is considered to fit the complex spectra of NGC 1068 in Section~\ref{sec:chandra_data}, were additional \chandra data are also included in the fit.

\begin{deluxetable}{lcccc}
    \tablewidth{0pt}
    \tablecaption{Best-fit results for each \nicer observation of NGC 1068. The energy bin describes the energy range for which the source signal dominates the background. The observations with zero exposure are not included. The normalization is given in units of photons/keV/cm$^2$/s at 1$\,$keV.}
    \label{tab:fitting_results_NGC}
    \tablehead{ T$_{start}$  & E$_{\mathrm{range}}$  & Normalization & $\Gamma_{\rm X}$ & $\chi^2/\mathrm{DOF}$ \\
    (MJD) & keV & ph/keV/cm$^2$/s & - & -}
    \startdata
    60005.38 & 0.3-2.0 &$1.47^{+0.09}_{-0.09}\times10^{-3}$&$4.37^{+0.08}_{-0.07}$& 19.6/25 \\ 
    60153.28 & 0.25-2.0 &$3.10^{+0.02}_{-0.02}\times10^{-3}$&$3.34^{+0.06}_{-0.06}$& 25.7/27 \\ 
    60183.04 & 0.25-2.0 &$3.25^{+0.08}_{-0.08}\times10^{-3}$&$3.31^{+0.03}_{-0.03}$& 60.1/30 \\ 
    60215.24 & 0.25-2.0 &$2.66^{+0.09}_{-0.09}\times10^{-3}$&$3.38^{+0.04}_{-0.04}$& 77.8/30 \\ 
    60253.60 & 0.25-2.5 &$2.44^{+0.01}_{-0.01}\times10^{-3}$&$3.53^{+0.06}_{-0.06}$& 81.6/38 \\ 
    60283.23 & 0.25-2.5 &$3.47^{+0.09}_{-0.09}\times10^{-3}$&$3.24^{+0.03}_{-0.03}$& 69.8/37 \\ 
    60314.13 & 0.25-2.0 &$3.28^{+0.01}_{-0.01}\times10^{-3}$&$3.28^{+0.04}_{-0.04}$& 43.1/30 \\ 
    60343.04 & 0.25-2.5 &$3.09^{+0.08}_{-0.08}\times10^{-3}$&$3.35^{+0.03}_{-0.03}$& 74.7/37 \\ 
    \enddata
\end{deluxetable}

To study the behavior of the two sources over the \nicer observing period, we make a light curve and plot the flux ($E^2dN/dE$) in units of erg\,cm$^{-2}$\,s$^{-1}$. As the energy range cannot be different for individual light curve measurements, we used a set energy range of 0.3-2\,keV for all the observations. We then start with fitting all the \nicer observations for a particular source with a power law, while keeping the photon index tied and allowing the normalization to vary in each observation. The derived flux per observation from this joint fit {are shown in Section.~\ref{sec:gamma_ray_data_sed} along with \fermi observations. }
The derived combined best-fit spectral index ($\chi^2/$DOF) is $\Gamma_{\rm X}$=$2.07\pm0.02$($1487.7/491$) for PKS 1502+106, and $\Gamma_{\rm X}$=$3.39\pm0.02$ ($980.5/233$) for NGC 1068. The resultant $\chi^2/DOF$ is higher for both sources using only \nicer data, which can be attributed to the fact that the activity levels of the observations are different with slightly different spectral shapes. To address this, we add joint observations with \nustar (PKS 1502+106) and \chandra (NGC 1068) to the corresponding \nicer observation which improve the fits. While the joint fitting for PKS 1502+106 makes use of a power-law model similar to the \nicer data fit (see Section~\ref{sec:nustar_x_ray_data_sed}), the joint fitting for NGC 1068 makes use of a more detailed model to account for the known spectral complexity of this target (see Section~\ref{sec:chandra_data}).


\begin{deluxetable}{lcccc}
    \tablewidth{0pt}
    \tablecaption{Best-fit results for each \nicer observation of PKS 1502+106. The energy bin describes the energy range for which the source signal dominates the background. The observations with zero exposure are not included. The normalization is given in units of photons/keV/cm$^2$/s at 1$\,$keV.}
    \label{tab:fitting_results_PKS}
    \tablehead{ T$_{start}$  & E$_{\mathrm{range}}$  & Normalization & $\Gamma_{\rm X}$ & $\chi^2/\mathrm{DOF}$ \\
    (MJD) & keV & ph/keV/cm$^2$/s & - & -}
    \startdata 
    \hline\hline 
    \multicolumn{5}{c}{\nicer Data}\\ 
    \hline\hline 
    60005.47 & 0.25-2.0 &$3.52^{+0.10}_{-0.10}\times10^{-4}$&$1.99^{+0.04}_{-0.04}$& 97.0/30 \\ 
    60008.32 & 0.25-2.0 &$3.63^{+0.03}_{-0.03}\times10^{-4}$&$1.87^{+0.15}_{-0.15}$& 25.8/26 \\ 
    60031.03 & 0.25-2.5 &$2.68^{+0.07}_{-0.07}\times10^{-4}$&$2.20^{+0.04}_{-0.04}$& 108.9/40 \\ 
    60032.65 & 0.25-2.0 &$2.99^{+0.04}_{-0.04}\times10^{-4}$&$2.30^{+0.20}_{-0.19}$& 26.2/21 \\ 
    60062.86 & 0.25-3.0 &$4.01^{+0.03}_{-0.03}\times10^{-4}$&$1.68^{+0.14}_{-0.14}$& 57.9/36 \\ 
    60063.06 & 0.25-4.0 &$4.90^{+0.01}_{-0.01}\times10^{-4}$&$1.30^{+0.05}_{-0.05}$& 267.1/55 \\ 
    60064.02 & 0.25-2.0 &$3.55^{+0.04}_{-0.04}\times10^{-4}$&$1.57^{+0.19}_{-0.20}$& 19.0/23 \\ 
    60093.06 & 0.3-2.0 &$3.39^{+0.02}_{-0.02}\times10^{-4}$&$2.96^{+0.09}_{-0.09}$& 175.7/25 \\ 
    60123.77 & 0.3-2.0 &$3.20^{+0.01}_{-0.01}\times10^{-4}$&$1.99^{+0.06}_{-0.06}$& 50.3/27 \\ 
    60124.04 & 0.25-2.0 &$2.95^{+0.03}_{-0.03}\times10^{-4}$&$2.02^{+0.15}_{-0.15}$& 49.9/23 \\ 
    60125.14 & 0.25-2.1 &$3.65^{+0.03}_{-0.03}\times10^{-4}$&$1.82^{+0.14}_{-0.14}$& 28.6/25 \\ 
    60158.38 & 0.25-2.1 &$4.84^{+0.02}_{-0.02}\times10^{-4}$&$1.51^{+0.08}_{-0.08}$& 33.9/28 \\ 
    60159.15 & 0.25-2.0 &$3.12^{+0.02}_{-0.02}\times10^{-4}$&$1.97^{+0.08}_{-0.08}$& 53.4/26 \\ 
    60170.06 & 0.25-2.0 &$3.18^{+0.01}_{-0.01}\times10^{-4}$&$1.92^{+0.06}_{-0.06}$& 57.6/29 \\ 
    60171.09 & 0.25-2.0 &$2.54^{+0.04}_{-0.03}\times10^{-4}$&$2.19^{+0.22}_{-0.22}$& 21.6/21 \\ 
    60313.44 & 0.25-2.0 &$2.82^{+0.02}_{-0.02}\times10^{-4}$&$2.22^{+0.09}_{-0.09}$& 53.8/27 \\ 
    60314.02 & 0.25-2.0 &$2.69^{+0.02}_{-0.02}\times10^{-4}$&$2.33^{+0.11}_{-0.11}$& 46.0/26 \\ 
    60315.50 & 0.25-2.0 &$2.94^{+0.01}_{-0.01}\times10^{-4}$&$2.20^{+0.06}_{-0.06}$& 62.5/29 \\ 
    60316.02 & 0.25-2.1 &$3.14^{+0.03}_{-0.03}\times10^{-4}$&$2.25^{+0.13}_{-0.12}$& 38.2/26 \\   \hline\hline  
    \multicolumn{5}{c}{\nustar Data}\\ 
    \hline\hline 
    60005.53 & 5.5-20.0 &$1.65^{+0.03}_{-0.01}\times10^{-4}$&$1.53^{+0.52}_{-0.48}$& 36.4/25 \\ 
    60170.01 & 5.0-15.0 &$1.28^{+0.02}_{-0.08}\times10^{-4}$&$1.51^{+0.45}_{-0.45}$& 22.4/21 \\ 
    60343.21 & 5.0-15.0 &$3.04^{+0.04}_{-0.02}\times10^{-4}$&$1.92^{+0.42}_{-0.41}$& 18.9/26 \\ 
    \enddata
\end{deluxetable}

%
%
%
%

\subsection{\nustar observations (only PKS 1502+106)}
\label{sec:nustar_x_ray_data_sed}
PKS 1502+106 has been observed three times with \nustar as part of our monitoring campaign: the observations, each $\sim$20\,ks long, were taken in March 2023 {(MJD:60005)}, August 2023 {(MJD:60170)}, and February 2024 {(MJD:60343)}. We reduced the \nustar observations, {which} were taken in SCIENCE mode, following the standard \nustar data reduction procedure. We used the \nustar Data 
Analysis Software (NUSTARDAS) v2.1.2 
to process the data coming from both the \nustar cameras, namely Focal Plane Module A (FPMA) and Module B (FPMB).
As the observations between the two modules were found to be relatively similar, we used the FPMA data for fitting purposes. We performed the reduction and calibration of the event files using the standard \texttt{nupipeline} script, using the response file from the CALDB v.20240104. In each observation, we extracted the source from a circular region centered at the optical position of the source, using a radius $r$$\sim$150$^{\prime\prime}$, derived via a visual inspection. This ensured that no bright object contaminated the emission from PKS 1502+106. We then extracted the background spectrum from an annular region with an inner radius of $r$$\sim$180$^{\prime\prime}$ and an outer radius of 
$r$$\sim$300$^{\prime\prime}$. We then derived the source and background spectra, as well as the corresponding ancillary response file (ARF) and redistribution matrix file (RMF) by using the  \texttt{nuproducts} tool.
The reduced spectra {are} grouped using the \texttt{grppha} tool from Xspec to ensure that there exist a minimum of 70 counts per bin, chosen to reduce the uncertainty per bin from the \nustar data. Similar to the \nicer fitting, the binned spectra are then fitted using a power law with Galactic absorption model ($TBabs*powerlaw$). The results of the fits are highlighted in Table~\ref{tab:fitting_results_PKS} (bottom three rows).

\begin{figure}[ht!]
   \begin{center}
   \begin{tabular}{c}
    \includegraphics[width=\linewidth]{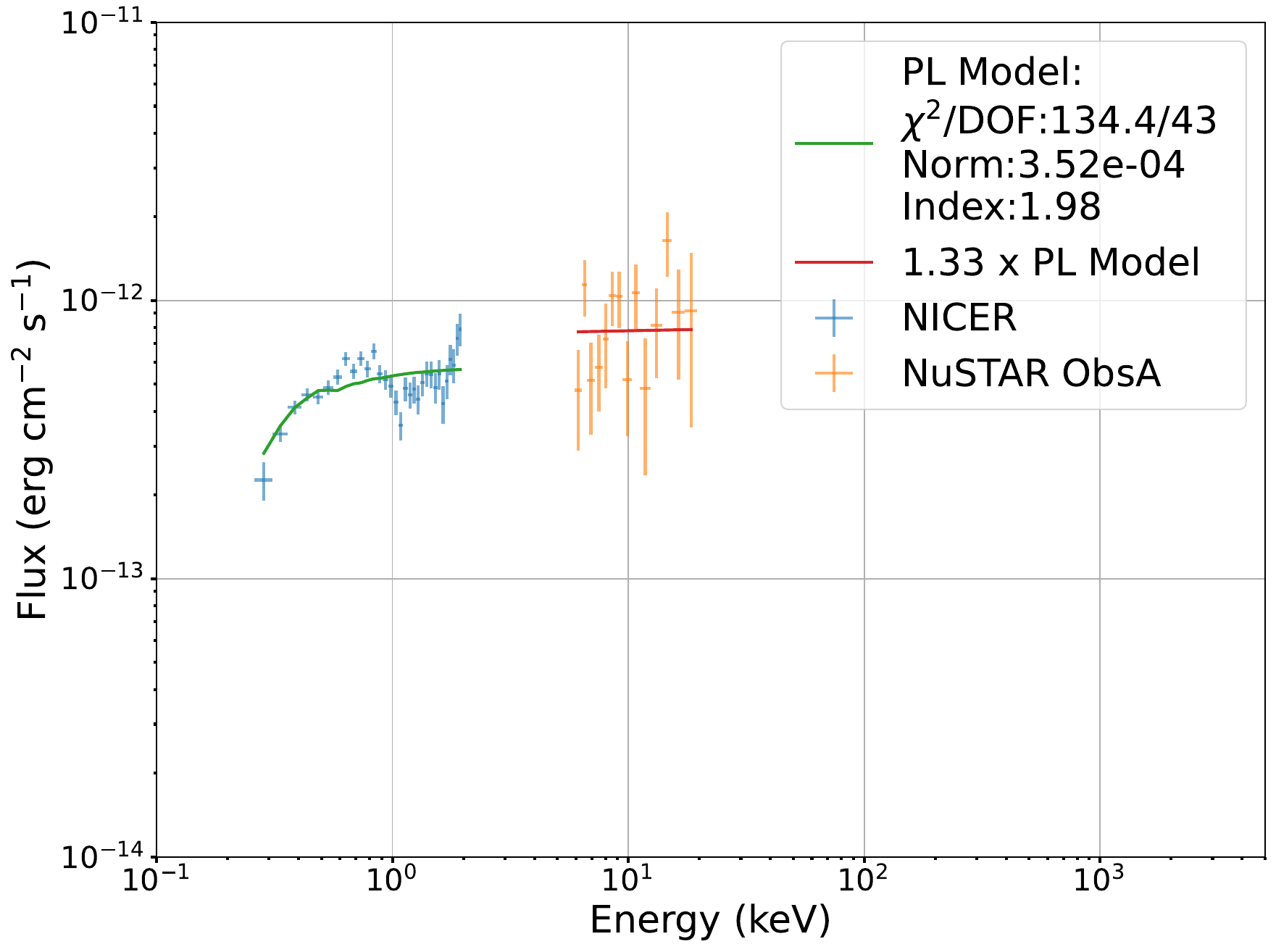} \\
    \includegraphics[width=\linewidth]{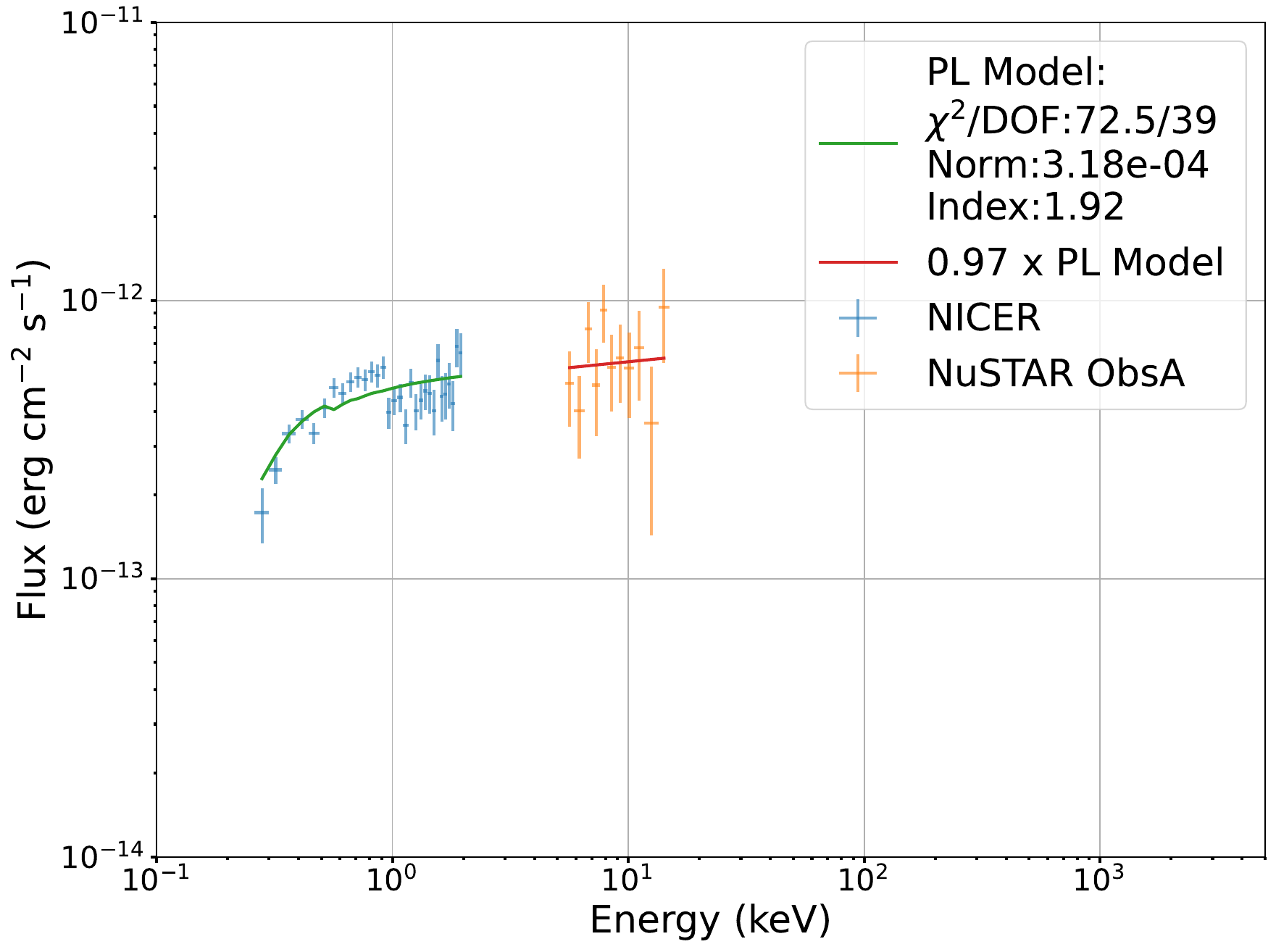} 
   \end{tabular}
   \end{center}
   \caption{Joint-fits for the \nicer+\nustar observations of PKS 1502+106 made on {MJD:60005} (top plot) and {MJD:60170} (bottom plot).
   }
  \label{fig:joint_fit_results}
\end{figure}

We also derived the combined best-fit spectral index for the \nustar observations (binned between 5-15\, keV) to be $\Gamma_{\rm X}$=$1.58\pm{0.32}$ (32.55/36). 
This combined \nustar best-fit spectral index for PKS 1502+106 differs with the combined estimate from \nicer of $2.08\pm0.01$. This can be attributed to the different energy ranges covered by the satellites, with no overlap after cutting out the spectral regions affected by background (see Table~\ref{tab:fitting_results_PKS}). 
In addition to these individual fits, a joint fit for the observations made with \nicer in parallel to \nustar was also conducted. Unfortunately, one of the three joint observations of PKS 1502+106, corresponding to the \nustar observation made on {MJD:60343}, had a zero exposure measurement from \textit{NICER}; so no joint fitting was done. For the other two joint observations, fitting was performed using a power-law model with a photoelectric absorption component (TBabs) along with a constant scaling factor. The fit model is adapted based on the analysis done by \cite{pks_xmm_swift_fermi_outburst} using \swift and \textit{XMM} data.
The model used is described as:

\begin{dmath}
    constant*TBabs*powerlaw
\end{dmath}
All the model parameters are tied together for the \nicer and \nustar data, while the constant scaling parameter is fixed at 1 for \nicer and free to vary for \textit{NuSTAR} (FPMA observation). This scaling parameter is included to make sure the same model is used for both \nicer and \nustar data, with an option to vary the normalization for \textit{NuSTAR}, to account for potential cross-instrument calibration offsets.
The following results were derived for the two joint fits: for data collected during MJD:$60005$, the best-fit spectral index is $1.94\pm0.04$ with a $\chi^2/ \mathrm{DOF}$ of $116.1/43$. For data collected during MJD:$60170$, the best-fit spectral index is $1.92\pm0.06$ with a $\chi^2/ \mathrm{DOF}$ of $65.8/39$. The results of the joint fits are also shown in Figure~\ref{fig:joint_fit_results}. The fitting results seen in this study have a much higher $\chi^2$ value as compared to \cite{pks_xmm_swift_fermi_outburst}. A reason for this discrepancy can be an additional feature seen in the \nicer data for this source below 1\,keV which was not seen by the \swift data used by \cite{pks_xmm_swift_fermi_outburst}.

\subsection{Chandra Archival Data (only for NGC 1068)}\label{sec:chandra_data}
We searched the X-ray archives for any available X-ray observation of PKS 1502+106 and NGC 1068  taken between March 2023 and March 2024, which is during the \nicer monitoring campaign. 
We found a \chandra observation with an ObsID of 29071, {taken on January fourth 2024 (MJD:60313.73)} with an exposure of 10.2 ks \citep{chandra_archival_obs_23}, coincident with one of our \nicer observations. This \chandra observation is added to the sample and processed jointly with the coincident \nicer observation with ObsId of 6696021102 and exposure of 10 ks.  
The \chandra observations are reduced using the \texttt{ciao-4.17} tool  following the procedure given by the data processing guideline for pointlike sources \footnote{https://cxc.cfa.harvard.edu/ciao/threads/pointlike/index.html}. A circular region of radius $\sim3''$ is used to extract the source counts. To derive the background, four circular regions were chosen with a radius of  $\sim5''-6''$ with no overlaps with the source or among each other.

\begin{figure}[ht!]
   \begin{center}
   \begin{tabular}{c}
    \includegraphics[width=0.9\linewidth]{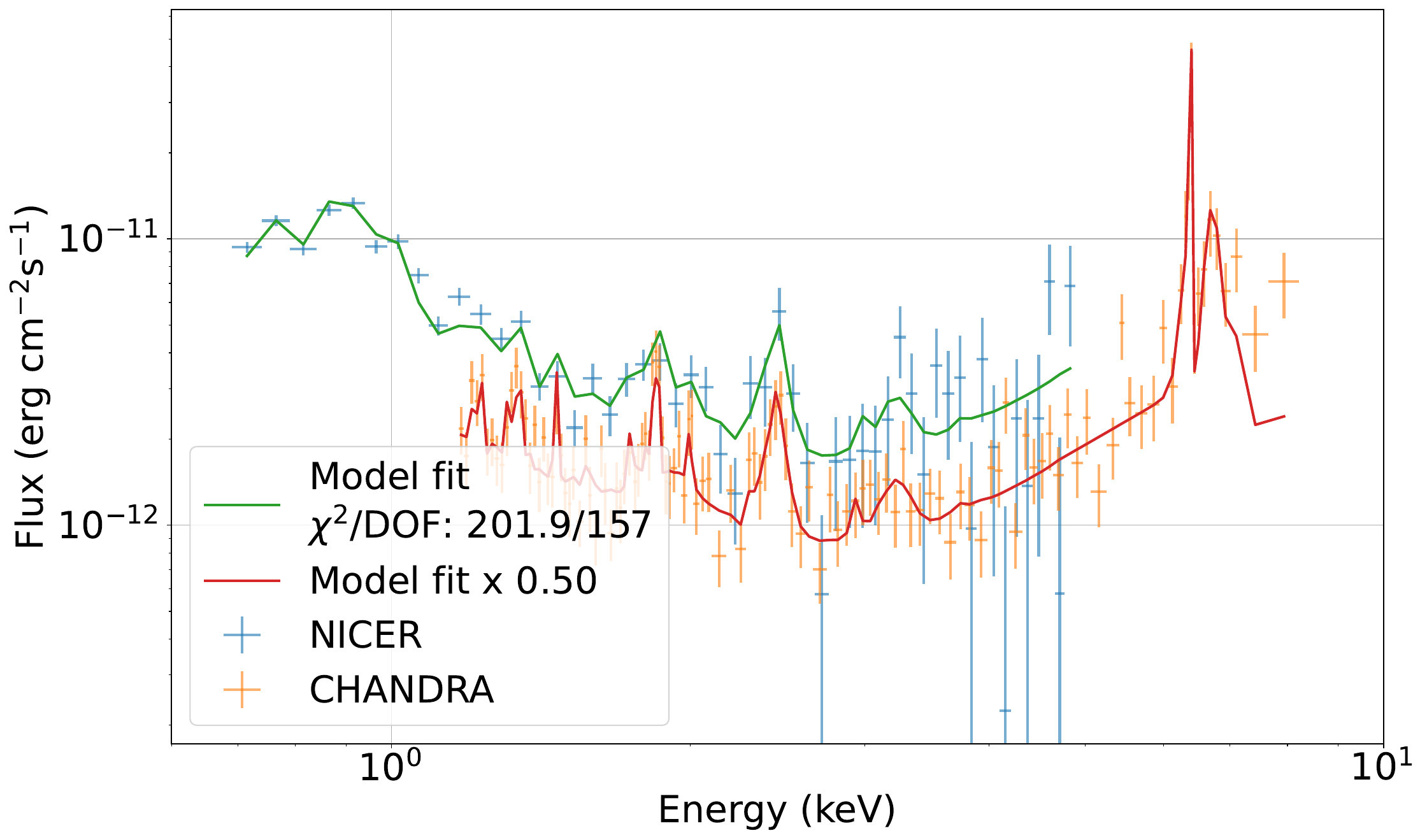} 
   \end{tabular}
   \end{center}
   \caption{Joint-fit for the \nicer+\chandra observation of NGC 1068 is shown here. The \nicer observation is made on {MJD:60314} and the \chandra observation was conducted on {MJD:60313}.}
  \label{fig:chandra_1068}
\end{figure}

Once the \chandra observations are reduced, we performed a joint \nicer-\chandra fit.  
\citet{1997MNRAS.289..443I} and \citet{1999MNRAS.310...10G} have highlighted the spectral complexity of NGC 1068, far exceeding that of typical Seyfert 2 galaxies. Its continuum includes at least three components: a thermal-like emission likely linked to starburst activity detected by ROSAT \citep{1992ApJ...391L..75W}, a cold reflection component, and a highly ionized reflection component \citep{1997MNRAS.289..443I, 1997A&A...325L..13M}. Numerous emission lines, mainly from ionized gas, are superimposed, although the neutral Fe K$\alpha$ line is also evident. Following the approach of \cite{indrani_ngc_1068}, we adopt an ionization parameter of $log\,\xi=0$ for the cold (neutral) reflector, and $log\,\xi$ kept free to vary for the warm (ionized) reflector.

Our adopted spectral model, following \citet{1999MNRAS.310...10G, 2001MNRAS.322..669B}, includes:
\begin{enumerate}
    \item a mekal thermal plasma,
    \item a cold reflection component  \citep[xillver;][]{xillver},
    \item an ionized reflection component \citep[\textit{xillver} with log$\xi$ free to vary;][]{indrani_ngc_1068},
    \item Several emission lines modeled using Gaussian profiles. Note that the Gaussian lines are only added when statistically required by the data.
\end{enumerate}

In XSPEC, the resultant final model is expressed as:
\begin{dmath}
Model =  const * phabs * (mekal + xillver_{cold} + xillver_{warm} + zgauss + zgauss + zgauss + zgauss)
\end{dmath}
Additional details of the fitting procedure and the relevant best-fit and emission line parameters are given in the Appendix. 
The model fit is shown in Figure~\ref{fig:chandra_1068}. The \nicer observations are reduced using the \texttt{3c50} background, but we also test this joint fit using the \texttt{scorpeon}\footnote{https://heasarc.gsfc.nasa.gov/docs/nicer/analysis_threads/scorpeon-overview/} background and find that the $\chi^2/\mathrm{DOF}$ is comparable using the two background models.

As discussed above, due to NGC 1068 being a heavily obscured AGN, the observed X-ray spectrum is particularly hard to characterize.
In particular, its 0.5--10\,keV emission is expected to be dominated by the reprocessed photons coming from the interaction between the intrinsic emission photons and the obscuring medium. The line-of-sight column density of the obscuring medium is N$_{\rm H,los}$$>$10$^{25}$\,cm$^{-2}$, which prevents a clear detection of the intrinsic emission \citep{Bauer_1068_obs,zaino20}. 
Nonetheless, the broad-band fit we performed allowed us to obtain an estimate of the intrinsic flux of the primary component, which varies in the range $\sim 3-4\times10^{-10}\,\mathrm{erg}\,\mathrm{cm}^{-2}\,\mathrm{s}^{-1}$ 
in the $2-10\,$keV energy range. This is then compared with the results of \cite{Bauer_1068_obs} to derive the 2-10 keV intrinsic flux from this work. We use a simple power law with photon index $\Gamma$=2.1 and 2--10\,keV luminosity $log(\mathrm{L}_{2-10})=10^{43.1}\mathrm{erg}\,\mathrm{s}^{-1}$, as reported in \cite{Bauer_1068_obs}.
This value is thus used as the intrinsic flux to compare the observed multi-wavelength and neutrino spectrum against the theoretical model of \cite{ngc1068_theory_model} in Section~\ref{sec:Comparision_to_models}.



%
%

\begin{figure}[ht!]
   \begin{center}
   \begin{tabular}{c}
    \includegraphics[width=\linewidth]{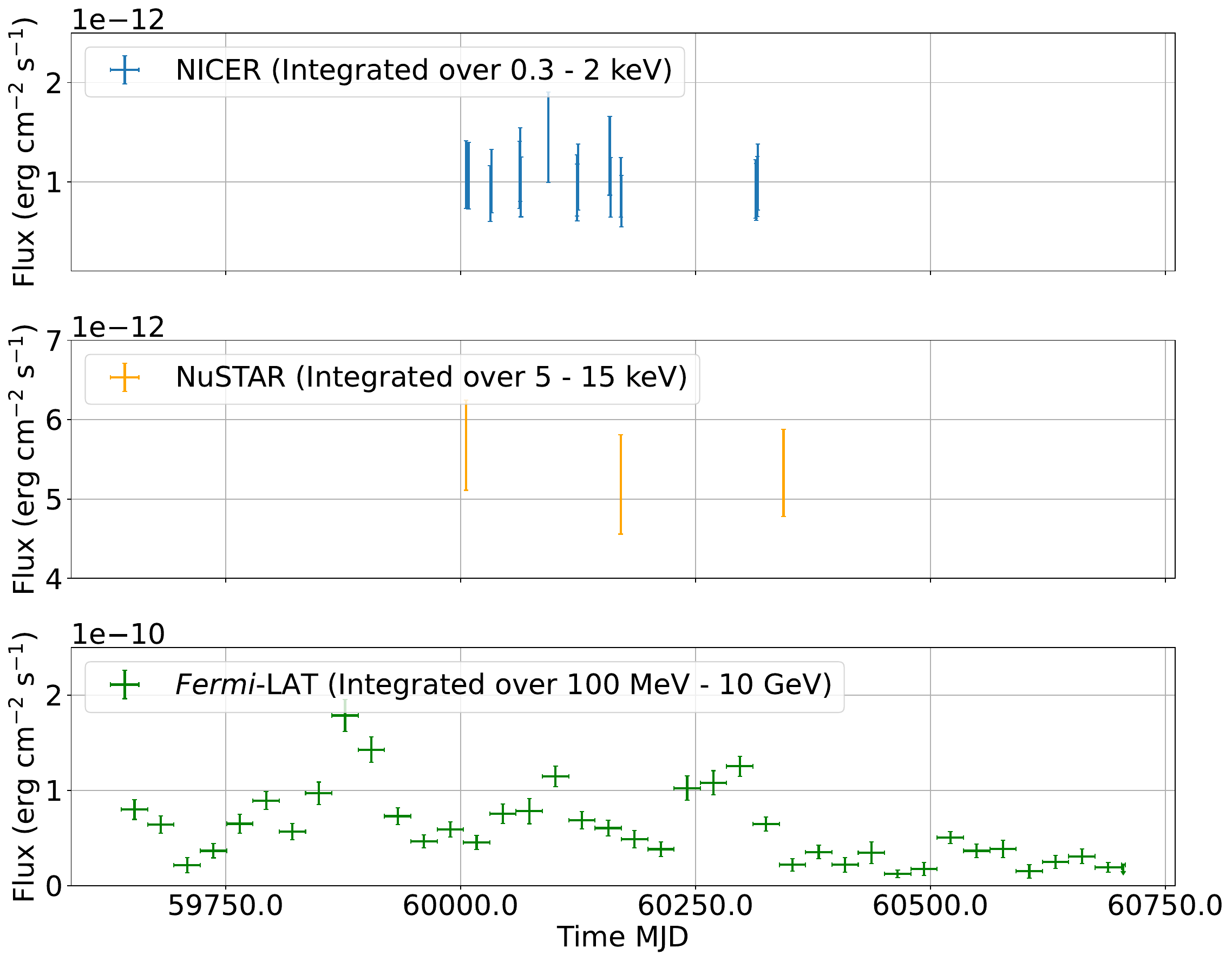}\\
   \end{tabular}
   \end{center}
   \caption{
   Light Curve for PKS 1502+106 derived using the data collected by \nicer (top panel),\nustar (middle panel) and \fermi (bottom panel) is shown here. 
   }
  \label{fig:PKS_lc_results}
\end{figure}

\begin{figure}[ht!]
   \begin{center}
   \begin{tabular}{c}
    \includegraphics[width=\linewidth]{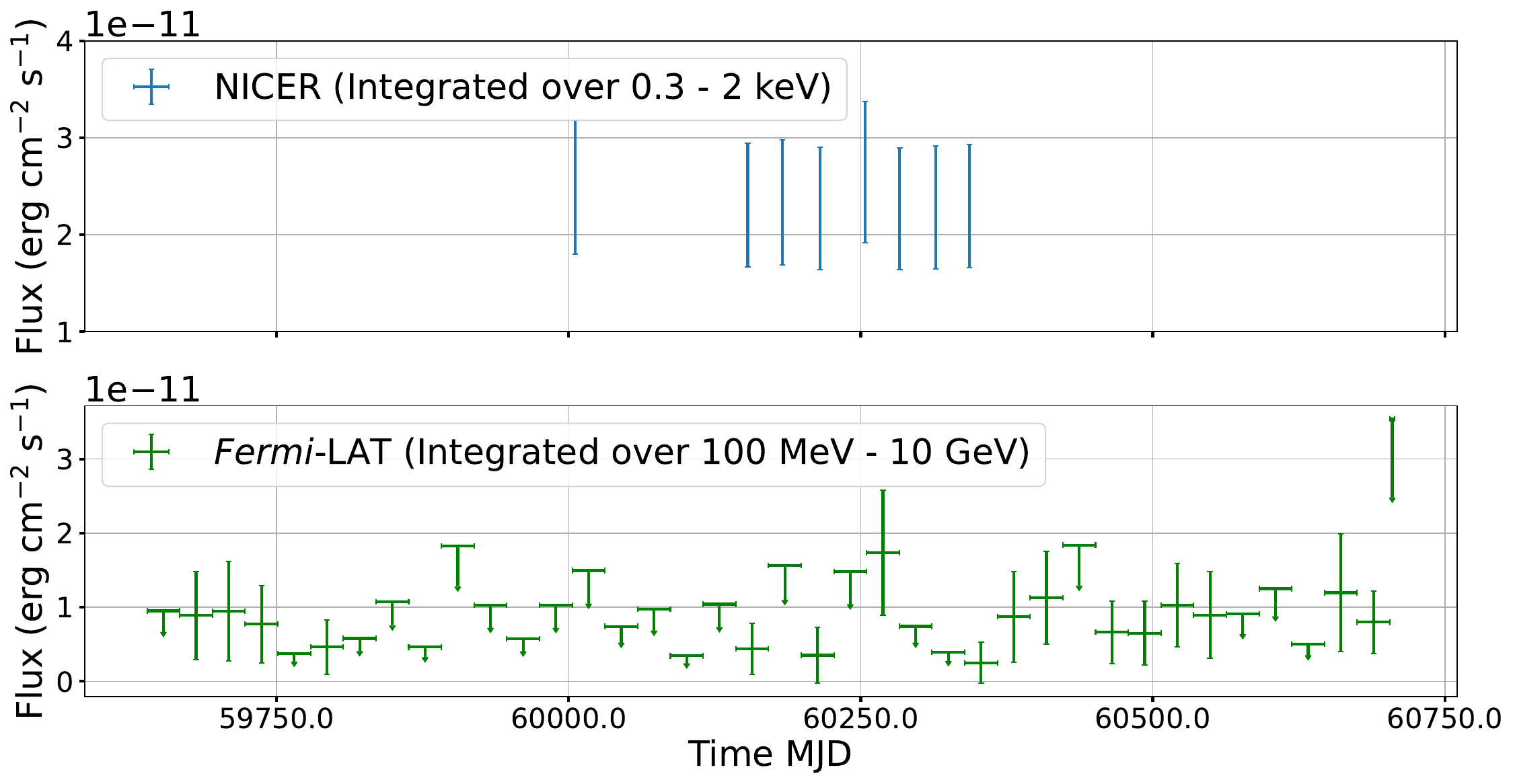}
   \end{tabular}
   \end{center}
   \caption{Light Curve for NGC 1068 derived using the data collected by \nicer (top panel) and \fermi (bottom panel) is shown here. The X-ray observations are taken between March 2023 (MJD:60004) to February 2024 (MJD:60341) while the {\it Fermi}-LAT light curve is derived using 4 week time bin periods from March 2022 (MJD:59639) to March 2025 (MJD:60735).
   }
  \label{fig:NGC_lc_results}
\end{figure}

\section{Gamma-ray data using \fermi}
\label{sec:gamma_ray_data_sed}

For both of the sources in our sample, we start by using 3 years of \textit{Fermi}-LAT data spanning from March 2022({MJD:59639}) to March 2025({MJD:60735}). 
The Pass-8 class data contains `SOURCE' photons that are binned over an energy range of 100\,MeV to 10\,GeV {and} are used for this study. The dataset is also filtered with a zenith angle of 90$^{\circ}.$ 
A region of interest (ROI) of 10$^{\circ}$ around the source position is used along with the data filters given by ``DATA\_QUAL>0 \&\& LAT\_CONFIG==1'' and an angular separation of 15$^{\circ}$ with the Sun. A sky model is then constructed comprising emission from background sources, diffuse Galactic emission and diffuse extragalactic emission\footnote{The  Galactic template gll$\_$iem$\_$v07.fits  and  the extragalactic template iso$\_$P8R3$\_$SOURCE$\_$V3$\_$v1.txt are used here and taken from https://fermi.gsfc.nasa.gov}. The background sources are found using the gamma-ray sources detected in the fourth {\it Fermi}-LAT DR3 catalog \citep{4fgl_dr3_paper} and located inside a 15$^{\circ}$ ROI around the source position, to account for photon leakage. A binned likelihood method was used for the study, along with the ``P8R3$\_$SOURCE$\_$V3'' LAT instrument response function.
Similar to the X-ray study, we first derive the {\it Fermi}-LAT light-curve for the two sources to check for any major flaring activity. As shown in Figures~\ref{fig:PKS_lc_results} and~\ref{fig:NGC_lc_results}, no significant flaring activity is seen for both of the sources over the 3 year period considered in this study.  

Two sets of SEDs for the {\it Fermi}-LAT data integrated over a period of 1 week and over a period of 3 years are then created. PKS 1502+106 shows some variation in the gamma-ray flux data {set} (see Figure~\ref{fig:PKS_lc_results}). So for each \nicer observation made on the time given in Table~\ref{tab:fitting_results_PKS}, a $\pm$3.5 day (giving a total of 1 week) \fermi observations are used to report the final SED. On the other hand, NGC 1068 is a Seyfert galaxy and not bright in gamma-rays {and} also had no significant variability seen in the generated light curve (Figure~\ref{fig:NGC_lc_results}). The {\it Fermi}-LAT data taken over a duration of 3 years is used to create the SED of NGC 1068. Both of the resultant SEDs {are discussed in more detail in }{Section~\ref{subsection:pks_neutrino}} (for PKS 1502+106) and {Section~\ref{subsec:Comparision_to_models_ngc}} (for NGC 1068).

\section{Comparing to multi-messenger model predictions}
\label{sec:Comparision_to_models}

\begin{figure*}[ht!]
   \begin{center}
   \begin{tabular}{c}
    \large MJD:60005 \\
    \includegraphics[scale=0.41]{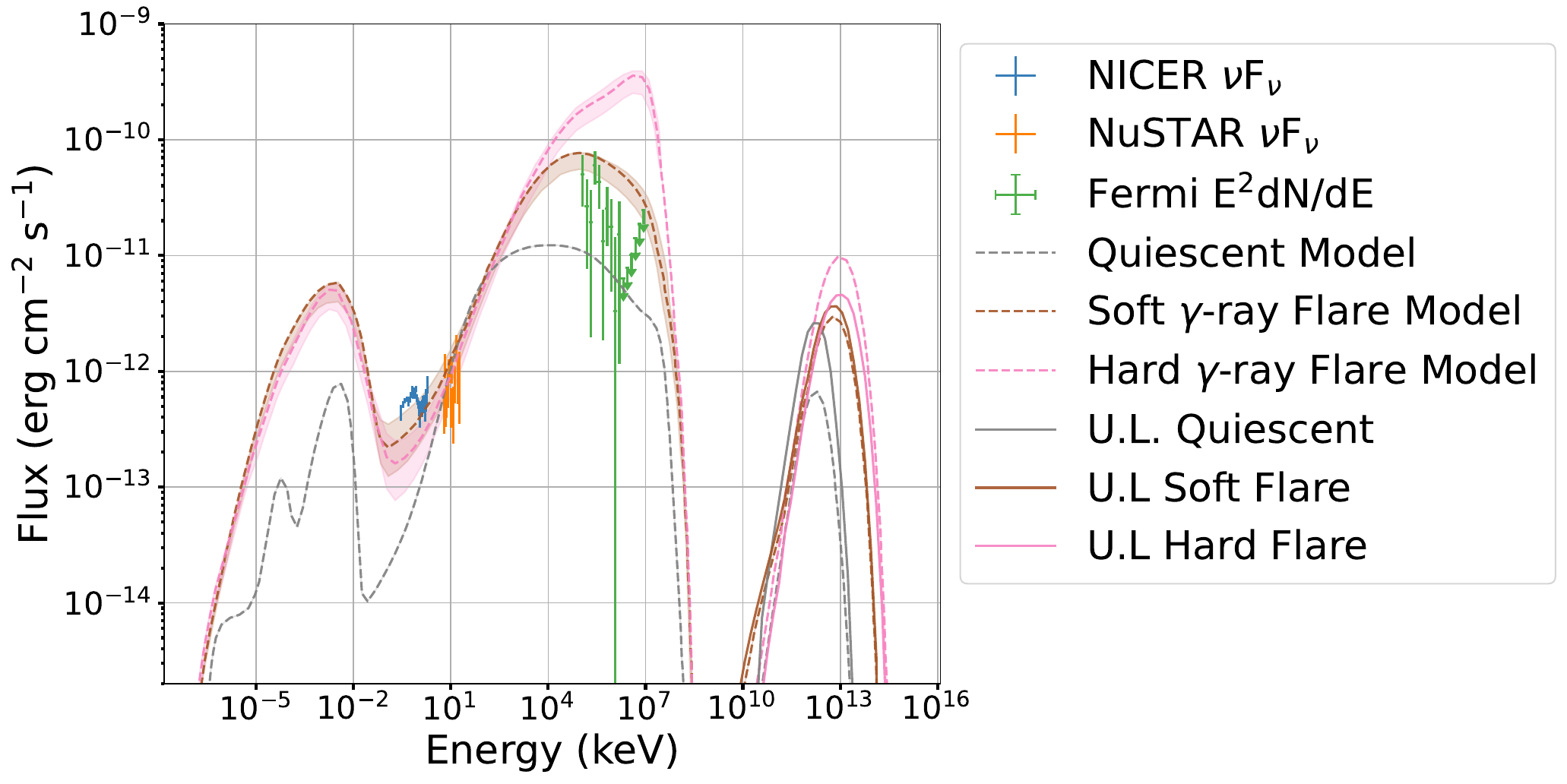}\\
    \large MJD:60170  \\
    \includegraphics[scale=0.41]{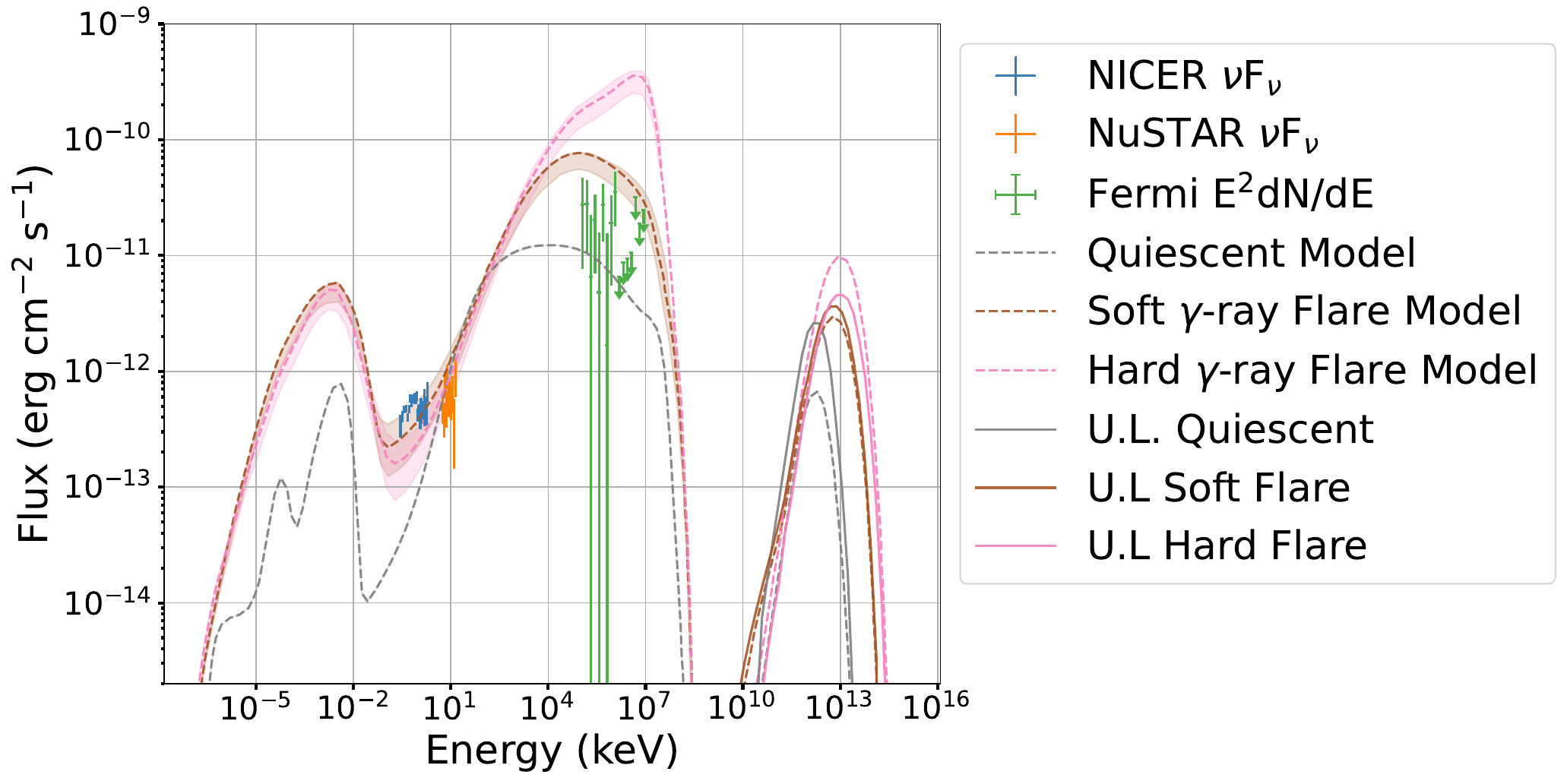}
   \end{tabular}
   \end{center}
  
   \caption{The multi-wavelength observations of PKS 1502+106 are shown (Top: {MJD:60005} and Bottom:{MJD:60170}, with the neutrino limits derived using \texttt{SkyLLH} and the theoretical models of \cite{Rodrigues_2021}. No fitting is done in this case and the models are shown only for comparison. The photon data includes  \nicer+\nustar+ 1-week \fermi observations.  
   }
  \label{fig:pks1502_with_model}

\end{figure*}

Before comparing the multi-messenger observations to model estimates, the neutrino flux measurements have to be derived.
To obtain the neutrino flux measurements and upper limits (in case of a non-detection), we make use of the \texttt{SkyLLH} tool described by \cite{skyllh_icrc_23}. This work makes use of the 10 years IceCube neutrino data \citep{data_icecube_10yr_data,icecube_10yr_data} sample which consists of track-like neutrino events. To test the \texttt{SkyLLH} setup, we first use a power-law spectral model as the source neutrino spectrum and compare it to the limits reported by \cite{10yrPStracks_tessa}. We use PKS 1502+106 as the test source and derive an upper limit of $6.1\times10^{-13}(\mathrm{TeV}^{-1}\mathrm{s}^{-1}\mathrm{cm}^{-2})$  which is compatible with {the} upper limit reported by \cite{10yrPStracks_tessa} of  $2.9\times10^{-13}(\mathrm{TeV}^{-1}\mathrm{s}^{-1}\mathrm{cm}^{-2})$. The discrepancy in the limits is a result of the reduced sensitivity of the public data analysis due to the detector response matrix binning, resulting in a slightly higher upper limit \citep[see also][for more details]{skyllh_icrc_23}.

\subsection{PKS 1502+106}
\label{subsection:pks_neutrino}
The blazar PKS 1502+106 has only been reported to have neutrino upper limits derived using a power-law source spectrum by IceCube \citep{10yrPStracks_tessa}. Studies like \cite{Rodrigues_2021} show that the predicted neutrino spectral shape can be different from a power-law, making a direct comparison of the existing limits with neutrino theoretical models difficult. To address this, we use \texttt{SkyLLH} to change the neutrino source model to match with the theoretical model, allowing us to put constraints on a particular model.
We use a likelihood setup similar to the one described in \cite{skyllh_icrc_23} along with the \textit{Minuit} minimizer. 
Additionally in our \texttt{SkyLLH} setup, the probability density functions (PDF) used are based on kernel density estimation (KDE). As described by \cite{skyllh_kde_icrc21} the KDE method is a nonparametric approach which improves the PDF construction of the point-source analysis improving the statistical power of the study.

For this work, we test the quiescent and flaring (Hard and Soft Flare) leptohadronic models reported by \cite{Rodrigues_2021} and use them as the injected source spectrum as opposed to a simple power law. While there exist additional multi-messenger models like \cite{Rodrigues24_models}, we choose the \cite{Rodrigues_2021} as the model to be tested here, due to the separately modeled quiescent and flaring curves. We change the injected neutrino source spectra in our setup from a simple {power-law} (PL) to a {Log Parabola} (LP)-like shape. The LP-like spectrum is similar to the estimation by \cite{Rodrigues_2021} and makes our likelihood dependent on the peak energy and normalization parameters for the neutrino flux distribution curve. Note that the shape of the spectrum is not fitted and only the peak energy and normalization can be varied in this test. For all three cases, we fix the peak energy of the distribution based on the model being tested. This ensures that in case of a non-detection, the reported upper limits can be used to put constraints on the theoretical model being tested.
  
No significant neutrino emission is seen when compared to the background for all three cases, so we derive the upper limits and show them as solid lines in Figure~\ref{fig:pks1502_with_model}. While the quiescent and Soft Flare neutrino model predictions of \cite{Rodrigues_2021} lie below our measured upper limits, the Hard Flare model can be ruled out based on the derived neutrino upper limits from our study.

\begin{deluxetable*}{lcccccc}
    \tablewidth{0pt} 
    \tablecaption{Best-fit neutrino results of NGC 1068 using the tracks dataset observation from \cite{data_icecube_10yr_data,icecube_10yr_data}. Three cases are tested based on the neutrino source model used. All the fits for Case 1-3 are derived using the \texttt{SkyLLH} software. The flux measurements shown for the power-law cases of the archival measurement from \cite{ngc1068_22} and Case 1 (derived in this work) are the best-fit neutrino flux at 1 TeV, while for the rest of the cases the flux measurement is the best fit value at the corresponding peak energy. Cases 2-5 shown here make use of the neutrino model prediction reported by \cite{ngc1068_theory_model} and \cite{Murase_2022_hidden_heart_AGN} as the source model as opposed to a simple power-law. The uncertainties for Case 1 are estimated using the Hessian matrix of the likelihood function. The uncertainties for Case 2-5 are derived using the $1\sigma$ contour assuming a $\chi^2$ distribution with the relevant degrees of freedom.}
    \label{tab:neutrino_fitting_results_NGC}
    \tablehead{ Case Tested  & Index  & Peak Energy  & Neutrino Flux & $-log_{10}p_{local}$  & TS$_{\mathrm{fit}}$  & D.O.F  \\
     &  & ($\mathrm{GeV}$)  & ($\mathrm{erg}\,\mathrm{cm}^{-2}\,\mathrm{s}^{-1}$) & (detection) & (fit) } 
    \startdata
    \cite{ngc1068_22}   & $3.2^{+0.2}_{-0.2}$  & -  & $8.01^{+2.4}_{-3.01}\times10^{-11}$ & 7.0 & - & 2  \\ 
    (Archival result for comparison)   &   &   &  &  &  \\ 
    \hline\hline  
    \multicolumn{7}{c}{This work:}\\ 
    \hline\hline 
    Case 1: Power-Law  & $3.2^{+0.3}_{-0.3}$ & -  & $5.25^{+1.3}_{-1.3}\times10^{-11}$ & 4.33 &  21.07 &  2  \\ 
    Case 2: Eichmann Model & - & $33.4$ & $1.01^{+0.27}_{-0.28}\times10^{-10}$ & 4.57 & 19.35  &  1 \\ 
    (Fixed peak energy) &  &  &  &  &  &  \\ 
    Case 3: Eichmann Model  & - & $3.4^{+48.8}_{-3.3}\,$ & $7.86^{+3.3}_{-2.9}\times10^{-10}$ & 4.27 &  $20.73$ &  2 \\ 
    (Varying peak energy)  &  &  &  &  &  &  \\ 
    Case 4: Murase Model & - & $822.5$ & $4.01^{+1.2}_{-1.1}\times10^{-11}$ & 4.63 & $18.64$  &  1 \\ 
    (Fixed peak energy) &  &  &  &  &  \\ 
    Case 5: Murase Model  & - & $110.2^{+652.4}_{-107.8}\,$ & $1.21^{+4.6}_{-5.1}\times10^{-10}$ & 3.96 &  20.93 &  2 \\ 
    (Varying peak energy)  &  &  &  &  &  &  \\
    \enddata
\end{deluxetable*}

\subsection{NGC 1068}
\label{subsec:Comparision_to_models_ngc}
The Seyfert galaxy, NGC 1068 has been detected by IceCube \citep{ngc1068_22} with a neutrino flux measurement that makes use of a power-law with a best-fit spectral index of $3.2\pm0.2$. There exist theoretical multi-messenger models \citep{ngc1068_theory_model,Murase_2022_hidden_heart_AGN} to explain this measured neutrino flux. These models make use of available multi-wavelength photon data along with the neutrino measurements to predict the neutrino flux spectrum. While the models use the IceCube power-law flux measurement to fit their spectrum, the shape of the neutrino spectrum follows a LP-like curve with a peak energy that varies based on the model presented. The neutrino model from \cite{ngc1068_theory_model} has a peak energy of $\sim 35\,$GeV. This peak energy changes to $\sim 822\,$GeV for the model presented by \cite{Murase_2022_hidden_heart_AGN}, for the $pp$ scenario with $\xi_B=0.01$, where $\xi_B$ is used to parameterize the magnetic field of the AGN. We use our \texttt{SkyLLH} setup to change the neutrino source spectrum from a simple power law to a LP-like curve following a spectral shape given by the \cite{ngc1068_theory_model} and \cite{Murase_2022_hidden_heart_AGN} models. The neutrino spectral shape is kept similar to the model tested and the only parameters fitted are the peak energy and the normalization making it a rigid horizontal shift or a scaled vertical shift of the model. 
This allows us to put updated constraints on the neutrino flux measurement and also allows us to test the model estimates with neutrino data.

We test five different cases for the source model. Case 1 makes use of a simple power-law model, which serves as a comparison of our setup with the results reported by \cite{ngc1068_22}. Case 2 and 3 substitute the power-law source model with a neutrino spectrum that has a fixed shape similar to the \cite{ngc1068_theory_model} neutrino model (dubbed as "Eichmann model" from this point). In Case 2, we fix the peak energy parameter, while in Case 3 the peak energy parameter is allowed to vary. Similarly, Case 4 and 5 make use of a neutrino model spectrum which follows a shape similar to the \cite{Murase_2022_hidden_heart_AGN} model, estimated for the minimal $pp$ scenario with $\xi_B=0.01$, where $\xi_B$ is used to parameterize the magnetic field of the AGN  (dubbed as "Murase model" from this point). Similar to the previous case, Case 4 fixes the peak energy parameter to the model estimate while Case 5 allows the peak energy parameter to vary.
For each of the cases, the local p-value denoted as $-log_{10}p_{local}$, which serves as the detection of the source as compared to the background, is computed. Note that the background is computed based on the spectral information and the reported {Test Statistic} ($TS$) and $-log_{10}p_{local}$ do not exhibit a monotonic relationship due to varying degrees of freedom and spectral shapes across cases.
Finally the best-fit parameter values are computed along with uncertainties and the best-fit TS and reported along with $-log_{10}p_{local}$ in Table~\ref{tab:neutrino_fitting_results_NGC}. 

We first compare the results reported by \cite{ngc1068_22} to the flux we derive using a power law neutrino source model (denoted as Case 1). We find that our best fit spectral index matches almost exactly with \cite{ngc1068_22}, while the flux is found to be compatible within uncertainties. The difference in the flux normalization 
values can be attributed to the difference in the neutrino data samples used for the two studies and the detector response matrix binning for \texttt{SkyLLH}. 
We then move to Case 2-5 to find that the derived $-log_{10}p_{local}$ does not vary significantly for these tested cases. The best-fit TS changes to account for the additional degree of freedom in the case of Case 3 and Case 5. While the fixed peak energy cases (Cases 2 and Case 4) are useful to provide normalization limits on the tested models, the free peak energy tests help us understand if the tested model is preferred by the neutrino data.

\begin{figure}[ht!]
   \begin{center}
   \begin{tabular}{cc}
    
    \normalsize Case 3 Eichmann Model \\
    \includegraphics[scale=0.4]{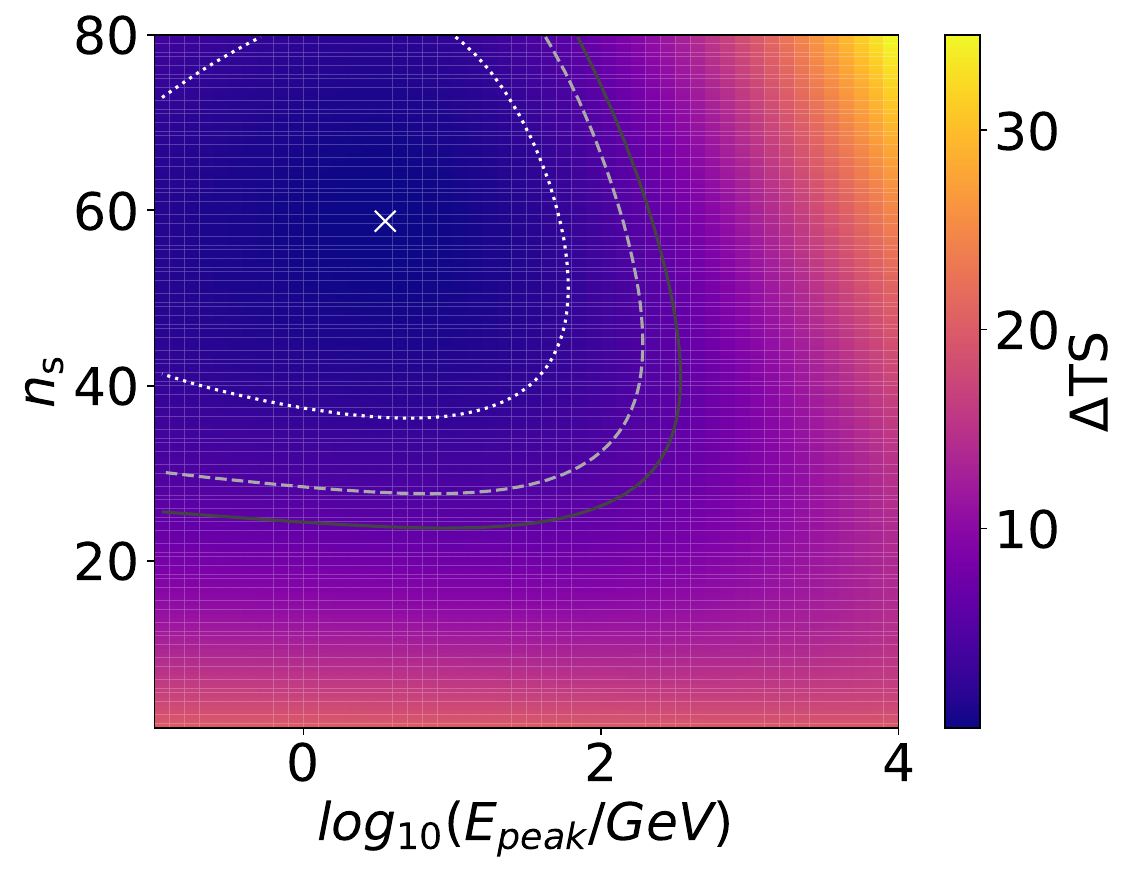}
    \\
    \\
    \normalsize Case 5 Murase Model \\
    \includegraphics[scale=0.4]{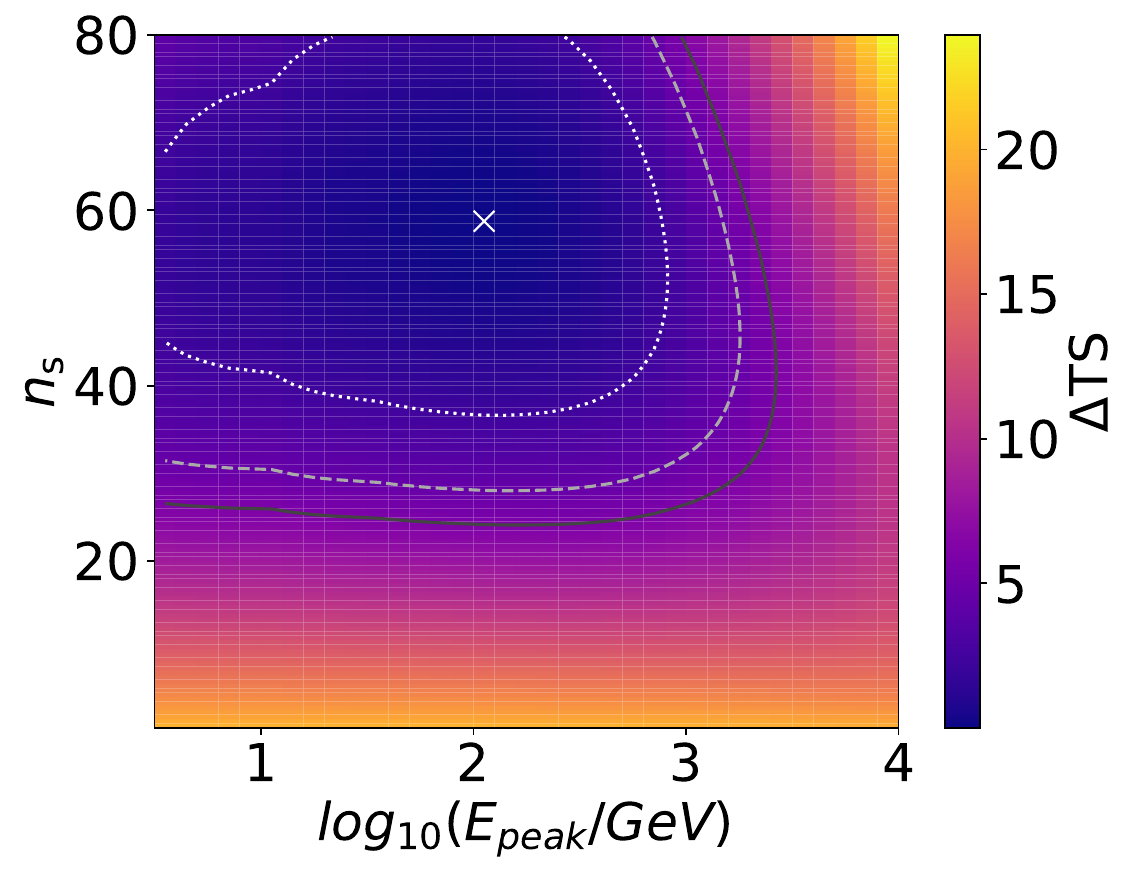} 
   \end{tabular}
   \end{center}

   \caption{The 2D likelihood profile of $E_{peak}$ and $n_S$ for Case 3 and Case 5 is shown here where the $\Delta TS = TS_{best fit} - TS(E_{peak},n_S)$ is derived for each combination of $E_{peak}$ and $n_S$. The white (dotted), gray (dashed) and black (solid) lines show the $68\%, 90\%$ and $95\%$ contour region respectively.
   }
  \label{fig:2d_likelihood}
\end{figure}

\begin{figure*}[ht!]
   \begin{center}
   \begin{tabular}{cc}
    \includegraphics[scale=0.32]{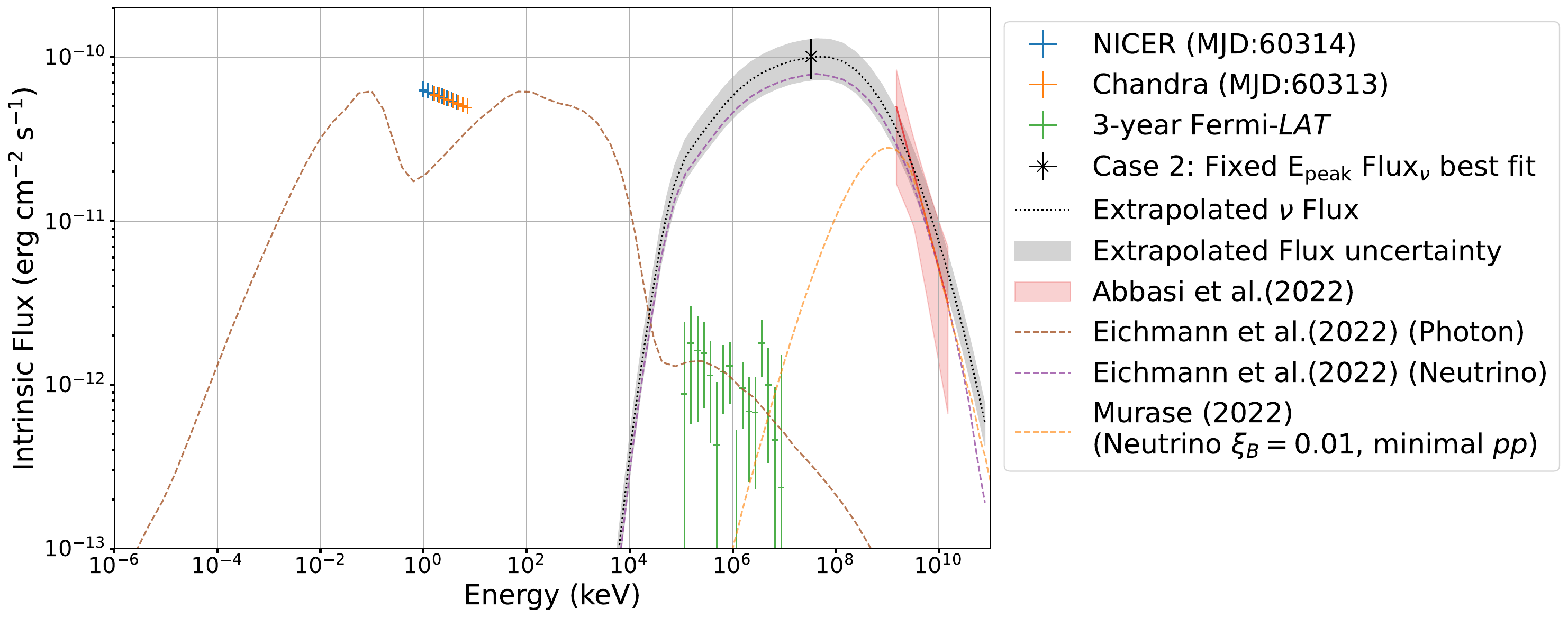}
    \\
    \includegraphics[scale=0.32]{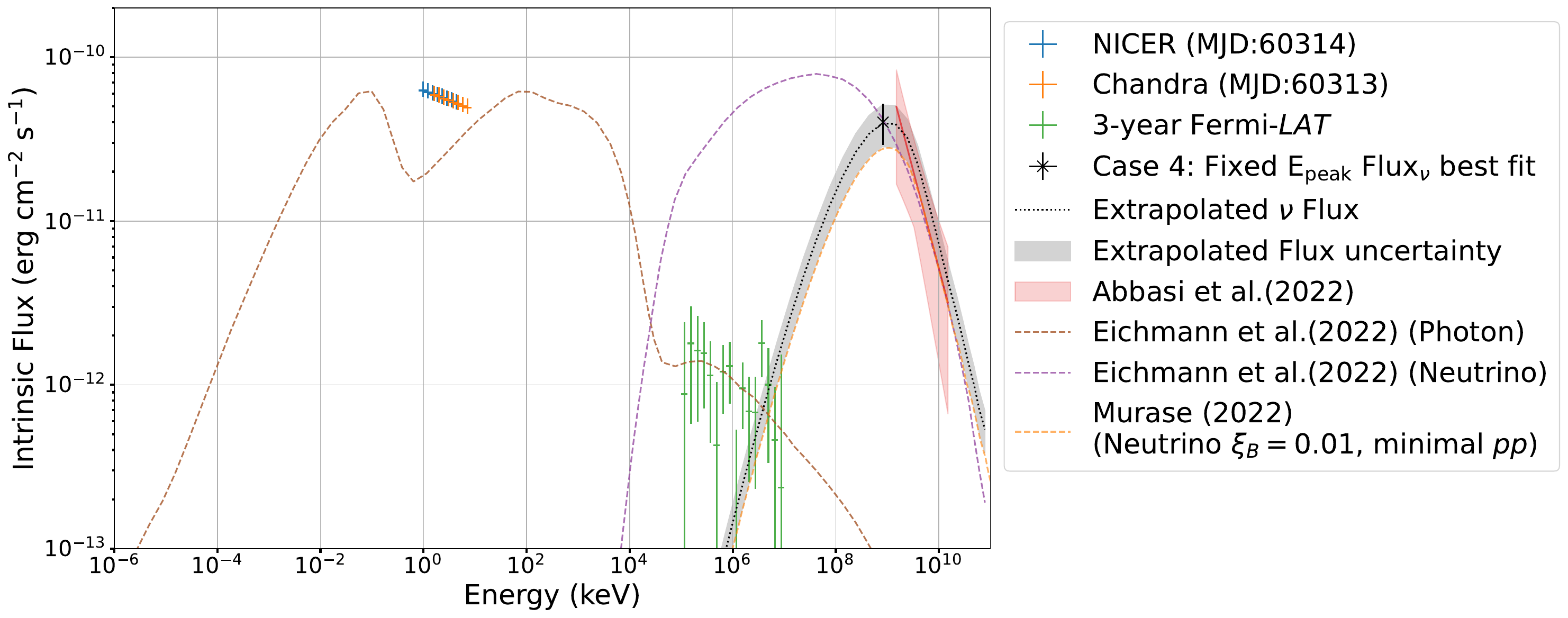} 
    \\
    \includegraphics[scale=0.32]{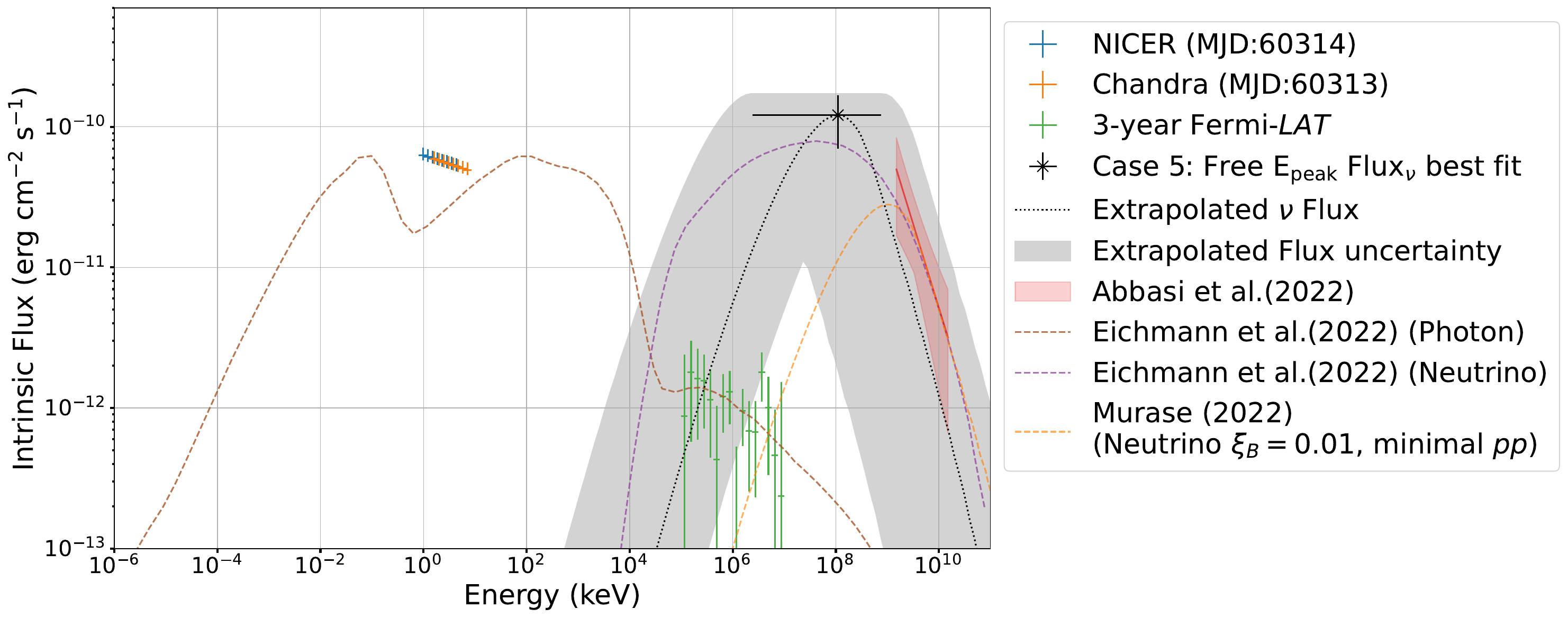}
   \end{tabular}
   \end{center}

   \caption{The multi-wavelength and multi-messenger observations for NGC 1068 are shown with neutrino measurements from \cite{ngc1068_22} (shown by the red shaded region). The observations shown here and derived using this study include: 1. The \nicer+\chandra data derived from Section~\ref{sec:chandra_data}, shown using the blue and orange data points respectively; 2. The \fermi observations are shown as green data points; and 3. The neutrino best-fit measurements, for three different neutrino source spectrum cases and derived using \texttt{SkyLLH} (see Table~\ref{tab:neutrino_fitting_results_NGC}) is shown as a black data point for each case (Case 2: Top, Case 4: Middle and Case 5: Bottom Plot). The extrapolated flux over 10\,GeV to 10\,TeV and the relevant uncertainty is also shown by a black dotted line with a gray shaded region. 
   The theoretical model predictions from \cite{ngc1068_theory_model} and \cite{Murase_2022_hidden_heart_AGN} also shown using dashed lines just for comparison. Note that no fitting is performed of the observed photon data (data points and solid black line) with the theoretical model (dashed lines).  
   }
  \label{fig:ngc1068_with_model}
\end{figure*}

We start with Case 3, which uses the Eichmann model with the peak energy allowed to vary. We derive a flux measurement which is a factor of 10 higher than the model and a best fit peak energy of 3.4\,GeV which is outside the sensitivity range of the Point Source (PS) tracks neutrino data sample \citep[$\sim100\,GeV-1\,PeV$; see][]{icecube_10yr_data}. Due to this, while the best-fit results from Case 3 are reported in Table~\ref{tab:neutrino_fitting_results_NGC}, they are not used to extrapolate a flux similar to Case 2 (see Figure~\ref{fig:ngc1068_with_model}; top). However, we can still use the best-fit results from Table~\ref{tab:neutrino_fitting_results_NGC} to understand why a lower best-fit energy is preferred by the model.
This case has two degrees of freedom, peak energy ($E_{peak}$) in GeV and normalization derived by fitting the number of neutrino signal events ($n_s$). We use different combinations of $E_{peak}$ and $n_s$ to derive a likelihood contour profile. A $\Delta TS (E_{peak},n_S)$ value is derived for each combination of $E_{peak}$ and $n_S$ using $\Delta TS = TS_{best fit} - TS(E_{peak},n_S)$, where the  TS values are derived by the using the corresponding likelihood ratios found by subtracting the likelihood at the point tested from the null likelihood. The resultant likelihood profile is shown in Figure~\ref{fig:2d_likelihood} (top). Additional lines denoting $68\%$, $90\%$ and $95\%$ quantiles for a $\chi^2$ distributions with 2 degrees of freedom are also added assuming Wilk's theorem \cite{wilks_thm}. It is seen from the profile that the fit prefers an energy with maximum upper limit uncertainty of $log_{10}(1.7)$ or $\sim$50 GeV which is already lower than sensitivity range of the PS tracks neutrino data sample. As the model tested has a rigid shape, a possible reason for this is that the neutrino observations are trying to fit the higher energy portion of the Eichmann model shifting the peak energy to lower values. Upon extrapolating the flux using the shape of the Eichmann model, it was seen that the flux measurement and shape of the spectrum match closely to the power-law flux reported by \cite{ngc1068_22}. 
However the resultant best fit flux at the peak energy of 3.4\,GeV for this Case  will require a proton luminosity that exceeds the bolometric luminosity. 
This implies that the Eichmann model is not soft enough to fit the neutrino data and a softer model is required. Note that this setup only shifts the tested model to find the best-fit peak Energy, while the shape of the model is rigid. The \texttt{SkyLLH} setup used in this work is not set to fit an additional parameter which can change the shape of the spectrum. The results of this Case also specify the need for a more detailed fit of the neutrino data with a more generic model to explain this result.
Future upgrades to the likelihood fitting along with future neutrino measurement at lower energies from IceCube detector upgrades \citep{icecube_gen2_20}, will also be able to better understand the flux behavior and update models similar to \cite{ngc1068_theory_model}.

Similar to Case 3, we test the \cite{Murase_2022_hidden_heart_AGN} model in Case 5 by changing the spectral model and freeing the peak energy and $n_S$. In this case the best fit energy lies within the sensitivity range of the PS tracks neutrino data sample $\sim100\,GeV-1\,PeV$ \citep[see][]{icecube_10yr_data}. A likelihood contour similar to Case 3 is also derived for this case and shown in Figure~\ref{fig:2d_likelihood} (bottom), which allows us to understand the results reported in Table~\ref{tab:neutrino_fitting_results_NGC}. The Murase model fit also prefers a lower peak energy, with the lower uncertainty limit falling below the peak energy reported by the model (and shown in Case 4). 
While the detection $-log_{10}p_{local}$ value for Case 5 is lower than Case 4, the TS$_{fit}$ for Case 5 is higher, with more degrees of freedom and a best fit peak-energy inside the sensitivity range, implying that Case 5 cannot be ruled out.
Thus, a flux extrapolation is derived for Case 5, along with Case 4, over a wider energy range and shown in Figure~\ref{fig:ngc1068_with_model} for comparison.

The best fit measurements for Case 2, Case 4 and Case 5 are used to obtain neutrino flux extrapolation measurements over a wider energy range. This is done using the shape of the neutrino model taken from \cite{ngc1068_theory_model} and \cite{Murase_2022_hidden_heart_AGN} and extrapolating over it using the best-fit measurements. {This} resultant neutrino flux extrapolation is shown in Figure~\ref{fig:ngc1068_with_model} as a black solid line with a gray uncertainty band. Note that in this work we do not derive a flux limit over the energy band where IceCube observations are most sensitive \citep[similar to][]{ngc1068_22}, but derive the flux over a wider energy range for better comparison with theoretical models. This is highlighted by the fact that the fluxes derived in this work for NGC 1068 match with the ones reported by \cite{ngc1068_22} in the region where IceCube observations are most sensitive (highlighted by the red shaded region in Figure~\ref{fig:ngc1068_with_model}).

\section{Summary and Discussion}
\label{sec:conclusion}

In this study we analyzed the photon emission from the blazar PKS 1502+106, and the Seyfert II galaxy NGC 1068, in the X-ray and gamma-ray regime, along with the neutrino emission seen by IceCube. We have compared these observed data with theoretical models while fitting the neutrino portion of the model. The results derived for the two sources are different and  discussed in more detail below.

\subsection{PKS 1502+106}
We compare the photon observations and derived neutrino limits for PKS 1502+106 with the model estimates from \cite{Rodrigues_2021} in Figure~\ref{fig:pks1502_with_model}. Based on only the neutrino upper limits, we see that the modeled neutrino emission is in excess for the Hard Flare case, while the modeled emission for the Soft Flare and quiescent case is consistent with our derived upper limits. While the neutrino upper limits can help rule out the neutrino emission derived using the hard-flare model, the neutrino upper limits alone cannot distinguish between the quiescent and flaring cases. 
The lightcurves in Figure~\ref{fig:PKS_lc_results} suggest that while the gamma-ray emission of the source was fluctuating over a $\sim$3 year period considered here, no significant flaring activity was seen in X-ray 
during the one-year observation period. This implies that the blazar was not flaring during the observation and the one-week gamma-ray observation used in Figure~\ref{fig:pks1502_with_model} coincides with a time when the source was in a quiescent state. Comparing the quiescent model with the \fermi and \nicer data, we find that the model under-predicts the observations. 
Our results suggest the need to refit the emission model after combining the \textit{NICER}, \nustar, \fermi observations along with neutrino upper-limits and update the quiescent model. 

\subsection{NGC 1068}
Comparing to PKS 1502+106, NGC 1068 is more significantly detected in the IceCube data \citep{ngc1068_22}, enabling a more detailed test of various theoretically motivated spectral models. As summarized in Table~\ref{tab:neutrino_fitting_results_NGC}, we tested LP-like neutrino models using the spectral shape provided by \cite{ngc1068_theory_model} or \cite{Murase_2022_hidden_heart_AGN}, where the peak energy of the spectrum is either fixed to the model estimate (Case 2,4) or allowed to vary (Case 3,5). In all cases except for Case 3 where the best-fit value of the peak energy falls outside the IceCube sensitivity range (see Section~\ref{subsec:Comparision_to_models_ngc}),  the derived flux extrapolations based on the best-fit neutrino flux agree with the measurement of \cite{ngc1068_22} within uncertainties (see Figure~\ref{fig:ngc1068_with_model}). For Case 2 and 4, where the peak energy is fixed, the model prediction lies below the derived neutrino flux but is still within uncertainties. In Case 5, where the peak energy is freed, the model of \cite{Murase_2022_hidden_heart_AGN} yields a large flux at 100 GeV, although the associated uncertainties are large.

These results highlight the need of additional constraints on the neutrino data at lower energies. Such constraints will reduce the uncertainties associated with the best-fit measurements and improve the significance of detections. Future upgrades to the IceCube neutrino detector \citep{icecube_gen2_20} or the IceCube ``deepcore" data would be able to help constrain the models in this lower energy region.

As NGC 1068 is an obscured AGN (see Section~\ref{sec:chandra_data}), the derived photon observations from \nicer and \textit{Chandra} for NGC 1068 are difficult to obtain. 
Based on the observational data collected and the joint photon model tested, intrinsic flux measurements are derived (in form of a simple power law) which can be then compared with the neutrino flux measurements. While the model estimates of \cite{ngc1068_theory_model}, shown in Figure~\ref{fig:ngc1068_with_model}, predict only the proton emission,  our \nicer+\textit{Chandra} observations are consistent with the background corona estimates shown as the uncertainty band in Figure 5 of \cite{ngc1068_theory_model}.
For this source, while photon measurements alone in the X-ray regime are not able to constrain the model, the neutrino data can help test the validity of the model. Future observations in the MeV regime by detectors like \amego can also help provide valuable model constraints for this source \citep[see also][]{ngc1068_marco_amego_fermi}.



Our results highlight the importance of combining neutrino measurements with photon observations and adopting a multi-component approach to accurately model the emission from NGC 1068.
They also demonstrate that sensitive observations in the MeV regime along with improved low-energy neutrino detections by future experiments will be crucial for constraining multi-messenger models of this source.



%
%
\section{Acknowledgements}
\label{sec:Acknowledgements}

The authors would like to thank Chiara Bellenghi and Tomas Kontrimas for helpful discussions relating to the \texttt{SkyLLH} software. The authors will like to thank Filippo D'Ammando and David J Thompson for helpful comments as part of the \textit{Fermi}-LAT collaboration internal review. 

The \nicer and \nustar data shown in this work makes use of the data collected by the \nicer and \nustar missions, and supported the \nicer GO Cycle 5 proposal run. The \chandra observations are based on archival measurements (DOI:10.25574/29071) hosted by the High Energy Astrophysics Science Archive Research Center. 

The \fermi data is provided by NASA Goddard Space Flight Center and supported by the \textit{Fermi-LAT} Collaboration. The \textit{Fermi} LAT Collaboration acknowledges generous ongoing support
from a number of agencies and institutes that have supported both the
development and the operation of the LAT as well as scientific data analysis.
These include the National Aeronautics and Space Administration and the
Department of Energy in the United States, the Commissariat \`a l'Energie Atomique
and the Centre National de la Recherche Scientifique / Institut National de Physique
Nucl\'eaire et de Physique des Particules in France, the Agenzia Spaziale Italiana
and the Istituto Nazionale di Fisica Nucleare in Italy, the Ministry of Education,
Culture, Sports, Science and Technology (MEXT), High Energy Accelerator Research
Organization (KEK) and Japan Aerospace Exploration Agency (JAXA) in Japan, and
the K.~A.~Wallenberg Foundation, the Swedish Research Council and the
Swedish National Space Board in Sweden.

Additional support for science analysis during the operations phase is gratefully 
acknowledged from the Istituto Nazionale di Astrofisica in Italy and the Centre 
National d'\'Etudes Spatiales in France. This work performed in part under DOE 
Contract DE-AC02-76SF00515.

The IceCube neutrino observations are based on 10 years of archival data, supported by the Wisconsin IceCube Particle Astrophysics Center and made available by the IceCube Collaboration and hosted by Harward Dataverse  (DOI:10.7910/DVN/VKL316).

A.D. was supported by an appointment to the NASA Postdoctoral Program at NASA Goddard Space Flight Center, administered by Oak Ridge Associated Universities under contract with NASA. 
A.D. and J.V acknowledge support from  \nicer Guest Observer Cycle 5 (Proposal 6196). J.V. acknowledges support from Vilas Associate funding from the Office of the Vice Chancellor for Research and Graduate Education at the University of Wisconsin-Madison and the National Science Foundation (PHY-2209445 and PHY-2013102). 
S.M. research activities on this project were carried out with contribution of the Next Generation EU funds within the National Recovery and Resilience Plan (PNRR), Mission 4 - Education and Research, Component 2 - From Research to Business (M4C2), Investment Line 3.1 - Strengthening and creation of Research Infrastructures, Project IR0000012 – ``CTA+ - Cherenkov Telescope Array Plus''.
K.F. acknowledges support from the National Science Foundation (PHY-2238916) and the Sloan Research Fellowship. This work was supported by a grant from the Simons Foundation (00001470, KF). J.T. acknowledges support from the National Science Foundation (PHY-2209445). S.H. acknowledges support from the National Science Foundation Graduate Research Fellowship Program (DGE-2039655).

%
%
\bibliographystyle{aasjournalv7}
\bibliography{references}

\appendix{}
\section{Joint X-ray Spectral Analysis of NGC 1068}
\label{sec:appendix}

A single reflector cannot account for the full 0.5–10 keV X-ray spectrum of NGC 1068, particularly for the coexistence of neutral and He-/H-like iron lines. As shown by \citet{1999MNRAS.310...10G, 2001MNRAS.322..669B}, at least three distinct reflectors are required: a cold, a warm, and a hot one. The cold, optically thick reflector can produce the OVII fluorescent line but requires a low ionization state, insufficient for the higher ionization lines. Conversely, the hot reflector, responsible for the He-/H-like Fe lines, is too ionized to produce lines from lighter elements. Therefore, a third, intermediate-ionization warm reflector is needed to explain features such as Mg, Si, and S lines.

As discussed in the main Section~\ref{fig:chandra_1068}, following the approach of \cite{indrani_ngc_1068}, we adopt an ionization parameter of $log\,\xi=0$ for the cold (neutral) reflector, and $log\,\xi$ kept free for the warm (ionized) reflector.
Our adopted spectral model includes a mekal thermal plasma, a cold reflection component  (\textit{xillver}), an ionized reflection component (\textit{xillver} with log$\xi$ kept free) and several emission lines modeled using Gaussian profiles added only when statistically required by the data.
In our analysis the model used in \texttt{XSPEC} is expressed as:

\begin{dmath}
Model =  const*phabs*(mekal+xillver_{cold}+xillver_{warm}+zgauss+zgauss+zgauss+zgauss)
\end{dmath}

Here, \textit{const} accounts for the cross-calibration between \nicer and \chandra, and \textit{phabs} models Galactic absorption with $N_H = 3.32 \times 10^{20}$ cm$^{-2}$, fixed to the value derived by \citet{2013MNRAS.431..394W}. 
All Gaussian lines had free energy and normalization parameters, with line width $\sigma$ fixed at 0.1 keV. The redshift was held at $z = 0.0038$ \citep{2010A&A...518A..10V}, and the inclination angle fixed at $i = 63^\circ$ \citep{2004A&A...414..155M}. 
The common parameters between the two reflectors such as the photon index ($\Gamma$), cut-off energy ($\rm{E_{cut}}$), the iron abundance ($\rm{AF_{e}}$) and the normalization are tied together. Following \cite{2004A&A...414..155M}, we fix $\rm{E_{cut}}$ to 500 keV.

The spectral components can be summarized as follows:
\begin{enumerate}
\item Cold reflector modeled with {\it xillver} 
\citep{2014ApJ...782...76G}, using $log\xi$ = 0 and $R=-1$ to produce neutral Fe K$\alpha$ ($\sim$6.4 keV) and K$\beta$ ($\sim$7.06 keV) lines.

\item Warm ionized reflector via {\it xillver} with $log\xi$ free to vary, representing Compton-scattered reflection from ionized gas. 
\item Thermal plasma emission described with {\it mekal}, with temperature and abundances free to vary.
\item Emission lines modeled using {\it zgauss} for Fe XXV ($\sim$ 6.7 keV), Ar XVII ($\sim$ 3.2 keV), S XV–Si XIV ($\sim$ 2.4 keV), and Si XIII–Si XIV ($\sim$ 1.9 keV).
\end{enumerate}

The best fit parameters for the continuum are given in Table~\ref{tab:chandra_fitting_results_NGC}. The emission line parameters are given in Table~\ref{tab:chandra_fitting_results_NGC1}.

\begin{table}[hbt!]
\caption{Best fit parameters of the joint \nicer and \chandra fit for NGC 1068 discussed in the Appendix.} \label{tab:chandra_fitting_results_NGC}
\begin{center}
  \begin{tabular}{ccccccr}
\hline
Model  & &  parameter  & & Value \\
\hline
{\it mekal} & & $\rm{kT_{e}}$ & & 0.75$^{+0.09}_{-0.03}$ \\
&& abundance && 0.10$^{+0.03}_{-0.04}$ \\
&& norm ($\times 10^{-2}$) && 1.71$^{+0.69}_{-0.30}$ \\
{\it xillver} && $\Gamma$ && 2.12$^{+0.17}_{-0.50}$ \\
&& {\it $log \xi$} 1.00$^{+0.51}_{-0.31}$ \\
&& $AF_{e}$ && 1.54$^{+1.01}_{-0.69}$ \\
&& norm ($\times 10^{-4}$) && 3.02$^{+0.42}_{-0.25}$ \\
&& $const$ && 0.51$^{+0.03}_{-0.03}$ \\
&& $\chi^{2}$/dof && 201.9/157 \\
\hline
\end{tabular}  
\end{center}
\end{table}

\begin{table}
\caption{Best fit line energies along with normalization derived for the joint \nicer and \chandra fit for NGC 1068. Here, the line energy (E) is in keV and the normalization ($N_{E}$) is in units of $10^{-5}$ photons keV$^{-1}$ cm$^{-2}$ s$^{-1}$} \label{tab:chandra_fitting_results_NGC1}
\begin{center}
\begin{tabular}{ccccccr}
\hline
Parameter  & &  line  & & Value \\
\hline
E1    & & Fe He-like K$\alpha$ & &  6.74$^{+0.08}_{-0.09}$     \\
$N_{E1}$ & &  & &  6.81$^{+2.44}_{-2.40}$  \\
E2  & &  Ar XVII  & & 3.24$^{+0.14}_{-0.15}$   \\
$N_{E2}$ & &    & & 1.46$^{+0.99}_{-0.95}$  \\
E3       & &  S XV - Si XIV & &  2.45$^{+0.03}_{-0.03}$   \\
$N_{E3}$ & &  & & 3.97$^{+1.61}_{-1.52}$ \\
E4       & &  Si XIII - Si XIV  & & 1.93$^{+0.06}_{-0.05}$   \\
$N_{E4}$ & &    & & 3.76$^{+1.74}_{-0.87}$  \\
\hline
\end{tabular}
\end{center}
\end{table}

\end{document}